\documentclass[12pt,a4paper]{amsart}

\usepackage[english]{babel}
\usepackage[cp1250]{inputenc}

\usepackage[T1]{fontenc}

\usepackage{charter}

\usepackage{newcent}

\usepackage[natbibapa]{apacite}
\usepackage{longtable,lipsum} 
\usepackage{enumerate}

\usepackage{color}

\usepackage{mathpple}
\usepackage{mathptmx}

\SetSymbolFont{letters}{bold}{OML}{cmm}{b}{it}
\usepackage{array}
\usepackage{calc}
\usepackage{bookman}
\usepackage{amsmath,amssymb,amsthm}
\usepackage{bm}
\usepackage{amsbsy}
\usepackage{paralist}
\usepackage{latexsym}

\usepackage{fancyhdr}
 \usepackage{mathrsfs}

\usepackage{amsthm}
\usepackage{multirow}
\usepackage{bigstrut}
\usepackage{graphicx}
\usepackage{caption}
\usepackage{subcaption}
\usepackage{tabulary}
\usepackage{caption} 
\makeatletter
\def\th@plain{%
  \thm@notefont{}
  \itshape 
}
\def\th@definition{%
  \thm@notefont{}
  \normalfont 
}
\makeatother
\theoremstyle{definition}

\usepackage{eucal}
 \usepackage{amsaddr}

\DeclareMathAlphabet{\mathpzc}{OT1}{pzc}{m}{it}
 \DeclareMathAlphabet{\mathcal}{OMS}{cmsy}{m}{n}
\newcommand{\dtheta}{\mathrm{d}\theta}

\newcommand{\dxij}{\mathrm{d}x_{i}^{\left(1 \right) }}
\newcommand{\dxid}{\mathrm{d}x_{i}^{\left(2 \right) }}

\newcommand{\dxt}{\mathrm{d}\mathbf{x}_{t}}

\newcommand{\dwt}{\mathrm{d}\mathbf{W}_{t}}

\newcolumntype{C}[1]{>{\centering\arraybackslash}m{#1}}
\makeatletter
\renewcommand{\email}[2][]{%
  \ifx\emails\@empty\relax\else{\g@addto@macro\emails{,\space}}\fi%
  \@ifnotempty{#1}{\g@addto@macro\emails{\textrm{(#1)}\space}}%
  \g@addto@macro\emails{#2}%
}
\makeatother

\begin{document}

\title{\textbf{Bayesian models in geographic profiling}}
\author{Jana Svobodová}
\address{Department of Mathematics and Statistics\\
 Masaryk University\\
 Kotlářská 2, 611 37 Brno\\
 janyfa@mail.muni.cz}

\maketitle

\markleft{\sectionname}

\bigskip

\bigskip

\bigskip

\begin{abstract}
We consider the problem of  geographic profiling and offer an approach to
choosing a suitable model for each offender. Based on the analysis of the examined
dataset, we divide offenders into several types with similar behavior. According to the spatial
distribution of the offender’s crime sites, each new criminal is  assigned to the corresponding group. Then we choose an appropriate model for the offender and using 
Bayesian methods we determine the posterior distribution for the criminal’s anchor point. Our models include directionality, similar to models of Mohler and Short (2012).
 Our approach also provides a way to incorporate two possible situations into
the model – when the criminal is a resident or a non-resident. We test this methodology on
a real data set of offenders from Baltimore County and compare the results with  Rossmo’s approach. Our approach leads to substantial improvement over Rossmo’s method, especially in the presence
of non-residents.
\end{abstract}

\bigskip

\bigskip

\noindent\textbf{Keywords:} Bayesian data analysis; distance decay function; anchor point; resident; non-resident.

\bigskip

\noindent\textbf{Running head:} Bayesian models in geographic profiling
\newpage

\section{Introduction}
The problem of geographic profiling aims at finding the common location of a criminal
(place of residence, workplace, favourite pub, etc.). Given the  knowledge of places where the criminal committed 
a series of crimes we want to estimate the so called anchor point $ \mathbf{z}=\left(z^{\left( 1\right) },z^{\left( 2\right) } \right)\in \mathbb{R}^{2} $.

There are several approaches to locate the anchor point. One approach is based on \textit{spatial distribution strategies} 
which estimate directly the anchor point by various methods. Such techniques include \textit{the centroid method},\textit{ center of minimum distance} or \textit{the circle method} \citep{canter}.

Another group of  techniques,  usually called probability distribution strategies, 
uses  \textit{hit score functions}. In order to construct such a function one has to choose a distance metric and a \textit{distance
decay function} which  distinguish individual methods in this group. The most popular ones include \textit{Rossmo's model CGT},
\textit{Canter's method} and \textit{Levine's method} \citep{canter,canter2,levine,oleary1,rossmo2}. The  hit score function indicates a prioritised search area. 

We can write  the hit score function for all   $ \mathbf{y}\in \mathbb{R}^{2}$ in the form

\begin{equation}\label{SkorFunkce}
S\left(\mathbf{y} \right) = \sum_{i=1}^{n}f\left(d\left(\mathbf{x}_{i},\mathbf{y} \right)  \right) \, ,
\end{equation}
where $ d\left( \mathbf{x}_{i} , \mathbf{y}  \right)  $ denotes a distance metric between points $\mathbf{x}_{i}  $ and $\mathbf{y}  $, the function $ f $ is a distance decay function and $ \mathbf{x}_{1},\mathbf{x}_{2},\ldots, \mathbf{x}_{n} $ denote  known crime sites corresponding to the given offender.

Formula \eqref{SkorFunkce} is subject to criticism, since it does not provide a probability density and does not include 
geographic features of the given region and other variables related to  the criminal's behaviour \citep{mohler2}.  
Several studies  point out an important connection between a series of crimes  and geography of the region \citep{brantingham,canter2,rossmo1}.
This is one reason why an appropriate tool to treat the problem is provided by Bayesian methods \citep{levine,oleary1,oleary5, mohler2}. 
They allow to implement information available before analyzing  data. Moreover, as a result we obtain a 
posterior which corresponds to a true probability distribution. 

The fundamental problem is to choose a suitable model which would best characterize the searched criminal. \textit{O`Leary} 
proposes several options in \citep{oleary1} or \citep{oleary2}.     So far there exists no universal model which would well describe the behaviour 
of an arbitrary criminal. On the other hand, \textit{Mohler} and \textit{Short} offer a more general model
which 
could cover a broad range of criminals for a suitable choice of parameters  \citep{mohler2}. The question remains which one to choose and  how to choose the parameters. 

It is obvious that every criminal has his  own particularities. However,  when we are at the stage of investigation, 
it is 
difficult to specify and determine what his style of behaviour is, hence his reasons for choosing the 
crime location. This paper offers an approach to deal with this problem. 

In Section \ref{2}, we introduce Bayesian approach to geographic profiling and show how we can incorporate the required parameters into the model. Section \ref{3} describes the data set that we use,  explains the necessity to distinguish between the coordinate systems and outlines a conversion of geographic coordinates recorded in the dataset to the plane coordinates, for the use the Euclidean metric in our calculation. In Section \ref{4}, we deal with different types of offenders in our data set, explain differences between resident and non-resident types of criminals and offer various models that are suitable for each type or subtype of offenders in our data set. Moreover, we suggest a method for dividing  offenders into the several categories based on the spatial distribution of their crime sites. Section \ref{5} describes how we can estimate the prior distribution for the used parameters. To obtain the parameters, we use kernel smoothing or logspline density estimation. In Section \ref{6}, we consider several cases of modelling with our data set. Then we illustrate the effectiveness of our methodologies  in comparison with  Rossmo's approach.

\section{Bayesian methods in geografic profiling}\label{2}
The choice of a model and suitable parameters for the offender's behaviour is one of the key parts of analysis.
The probability function (or density) $p$ captures our knowledge about the given offender. In other words, this
function suggests how the offender chooses the crime site. It represents our uncertainty and lack of information. 
From this point of view it is possible to use Bayesian approach (unlike the frequentist approach which could be only used 
if the offender acts randomly). 

Geographic profiling assumes that the crime site selection is influenced by the offender's anchor point $ \mathbf{z} $, hence  $ p $ will depend on $ \mathbf{z} $. For this approach, it is important that the offender has a single anchor point that is stable during the crime series. Denote $\boldsymbol{\theta}=\left\lbrace\theta_{1}, \theta_{2},\ldots,\theta_{k} \right\rbrace   $ the vector of $ k $ additional parameters which also influence the offender's behaviour, 
therefore $p$. It may include the average distance which the offender is willing to travel to 
the crime site, or a direction preferred by the offender (e.g. related to the transport infrastructure of the area),
the size of the buffer zone, etc. The choice of such parameters depends on the knowledge and experience of the analyst, 
and on the information available.

If we denote by $ \mathbf{x}_{1},\mathbf{x}_{2},\ldots, \mathbf{x}_{n} $ the known sites of a series of crimes, we can describe a model of the offender's
behaviour by a function $ p\left(\left\lbrace\mathbf{x}_{1},\mathbf{x}_{2},\ldots, \mathbf{x}_{n}\right\rbrace  | \mathbf{z},\boldsymbol{\theta}\right) $. It expresses the probability that the offender with a unique anchor point $ \mathbf{z} $ and with the given values of parameters $ \boldsymbol{\theta} $ commits crimes at the locations $ \mathbf{x}_{1},\mathbf{x}_{2},\ldots, \mathbf{x}_{n} $.

We would like to find the best way to estimate the anchor point $ \mathbf{z} $. One possible approach is to use the maximum likelihood method. For
criminalistic purposes, it  is not very convenient, since it
provides a single point estimate. In geographic profiling, we
would prefer to obtain an area, which contains the anchor point
with a high probability. This can be done using Bayesian analysis
\citep{bolstad,carlin,damien,robert,sivia}.

Using Bayes rule we obtain

\begin{equation}\label{BayesVetaKrim}
p\left(\mathbf{z},\boldsymbol{\theta}|\left\lbrace \mathbf{x}_{1},\mathbf{x}_{2},\ldots, \mathbf{x}_{n}\right\rbrace  \right) =\frac{p\left(\left\lbrace  \mathbf{x}_{1}, \mathbf{x}_{2},\ldots, \mathbf{x}_{n}\right\rbrace |\mathbf{z},\boldsymbol{\theta}\right)\cdot p\left( \mathbf{z},\boldsymbol{\theta}\right)}{p\left(\left\lbrace \mathbf{x}_{1},\mathbf{x}_{2},\ldots, \mathbf{x}_{n}\right\rbrace  \right) }\,.
\end{equation}

Let us recall that     $ p\left(\mathbf{z},\boldsymbol{\theta}|\left\lbrace \mathbf{x}_{1},\mathbf{x}_{2},\ldots, \mathbf{x}_{n}\right\rbrace \right) $ denotes a posterior distribution, $ p\left(\mathbf{z},\boldsymbol{\theta}\right) $ is a prior distribution, $ p\left(\left\lbrace  \mathbf{x}_{1}, \mathbf{x}_{2},\ldots, \mathbf{x}_{n}\right\rbrace |\mathbf{z},\boldsymbol{\theta}\right) $ denotes a likelihood function and the denominator $ p\left(\left\lbrace \mathbf{x}_{1},\mathbf{x}_{2},\ldots, \mathbf{x}_{n}\right\rbrace  \right) $ is called evidence. For our purpose, the evidence is a normalization constant. We can thus omit it from~\eqref{BayesVetaKrim} while replacing equality $ \left(= \right)  $ by proportionality $ \left( \varpropto\right)  $.

Since we are only interested in the probability distribution for $ \mathbf{z} $, we can remove the unnecessary parameters by integrating over
all possible values of $  \boldsymbol{\theta} $. In addition, if we assume 
the independence of the anchor point $ \mathbf{z} $ and further parameters   $ \boldsymbol{\theta} $ and mutual independence of the crime sites of the offender, we
obtain  

\begin{equation}\label{BayesUmerKrimMarg}
p\left(\mathbf{z}|\left\lbrace \mathbf{x}_{1},\mathbf{x}_{2},\ldots, \mathbf{x}_{n}\right\rbrace  \right) \varpropto \idotsint\limits_{M_{ \boldsymbol{\theta}}}p\left( \mathbf{x}_{1}|\mathbf{z}, \boldsymbol{\theta}\right)\cdot \ldots \cdot p\left( \mathbf{x}_{n}|\mathbf{z}, \boldsymbol{\theta}\right)\cdot h\left(\mathbf{z} \right)\cdot g\left(  \boldsymbol{\theta}\right)\,\dtheta_{1}\ldots\dtheta_{k}\,,
\end{equation}
where $ M_{ \boldsymbol{\theta}}\subseteq \mathbb{R}^{k} $ denotes the domain of integration, $h\left(\mathbf{z} \right)  $ is a prior corresponding to the anchor point $ \mathbf{z} $ and $g\left(\boldsymbol{\theta} \right)  $ is a prior distribution corresponding to the other parameters $ \boldsymbol{\theta} $.

There are numerous  possibilities for choosing $p\left( \mathbf{x}_{i}|\mathbf{z},\boldsymbol{\theta}\right)  $, or $ p\left(\left\lbrace  \mathbf{x}_{1}, \mathbf{x}_{2},\ldots, \mathbf{x}_{n}\right\rbrace |\mathbf{z},\boldsymbol{\theta}\right) $ in modelling the
offender's behavior. Also, we need to consider the most reasonable
choice of the prior and which parameters enter the prior. All this depends on available information, data,
geographic area under consideration and other factors. Hence, we
will analyze the data set which we use in
this study. After a careful description of this data set, we will
choose appropriate models and tools for analysis.

\section{The data set}\label{3}

In this study we have used freely available data  about 88 serial
criminals who committed crimes in Baltimore County in the time period 1993-1997. Data can be found at \url{http://www.icpsr.umich.edu/CrimeStat/download.html}. Most offenders  in the dataset committed more crime types. These included forty-five criminals who predominantly committed larceny, twelve who predominantly committed assaults, ten who committed mostly burglaries and the same number of offenders who committed mainly vehicle theft. There were nine criminals who were mostly robbers and the remaining two offenders cannot be categorized.

The number of crimes per criminal varied from three to thirty-three. In total, the data set contains 962 crimes. Each crime includes information about the identification number of the crime, the identifier of the offender, UCR code, the latitude and longitude of the crime site and  the latitude and longitude of the anchor point.

For the calculation of the distance and the choice of the distance metric, we have to distinguish between the geographic coordinate system and the projected coordinate system. The use of the Euclidean metric can lead to an easier calculation -- some functions can be expressed analytically. Additionally, all crime sites and anchor points of all offenders from the data set are located only in an area of $72{.}5 \,\text{km} \times 48{.}7\, \text{km}  $. The distances between a resident's crime site and his anchor point are  mostly a few kilometers. Hence, the choice of a coordinate system, or the choice of a distance metric should not lead to significantly different results.

Coordinates in the data set are recorded in the geographic coordinate system, thus,  we have to convert them into the plane coordinates in order to use the Euclidean metric. We work with the most common projected coordinate system UTM (\textit{Universal Transverse Mercator coordinate system}) that divides the Earth between $ 84° $ S and $ 80° $ N latitude into 60 zones, each $ 6° $ of longitude in width  \citep{kennedy}. To use the Euclidean metric  all investigated points have to lie in the same zone. Our data set is compliant for using this system because all points are located in the zone 18. To transfer coordinates, we use a simplified version of relations which \textit{Johann Heinrich Louis Krüger } derived in 1912  \citep{kawase1,kawase2}.

\section{Procedures and models}\label{SecModData} \label{4}
When analyzing the data set, we will generally assume  the fact that the crime series was 
committed by the same offender as known. This assumption is realistic in practice, since investigators may tell that the crimes
are related to one person, 
according to the way the crime was committed, DNA analysis or other signs and evidence; they just do not know which one and where to find him.

Different types of offenders require different models, hence we need to analyze the character of offenders in
the given data set, and the way they choose the crime site. 

\subsection{Models for residents}\label{PodsecMarr}
\begin{figure}[h]
    \begin{center}
\includegraphics [trim= 6mm 5mm 2mm  5mm,width=\textwidth]{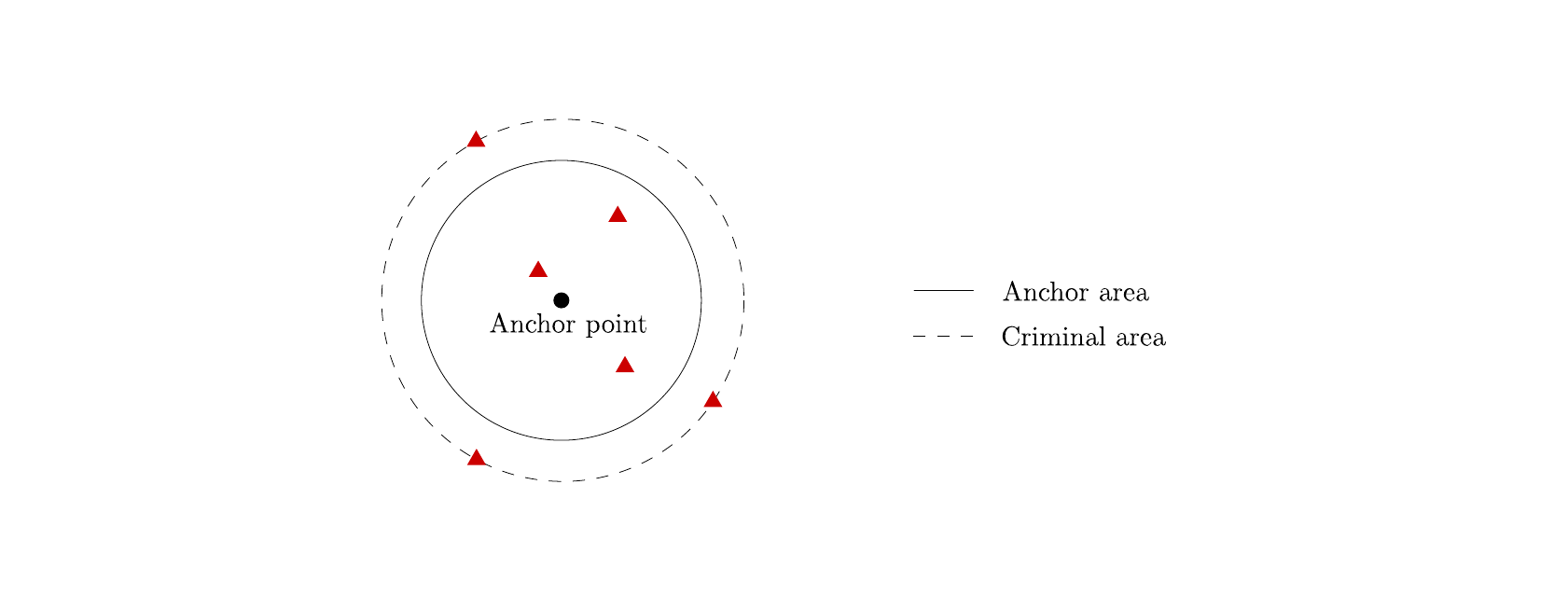}
\caption{\textit{Anchor and criminal areas of the resident offenders -- the dashed line circle  denotes the criminal area, the solid line circle is the anchor area, red triangles indicate crime sites, the black circle is the anchor point of the offender.}}
      \label{marauders}
    \end{center}
  \end{figure}

\index{Marodér}
First we will consider resident offenders. They commit crimes near their anchor point. Hence their 
criminal and anchor regions overlap, at least to a large extent. (see Fig. \ref{marauders}). 

After analyzing the data set, we can further classify offenders  into two subtypes, each of them requiring a 
different model.

\begin{figure}[h]
    \begin{center}
    \begin{subfigure}[b]{0.49\textwidth}
                 \includegraphics[trim= 0mm 0mm 0mm 0mm,width=\textwidth]{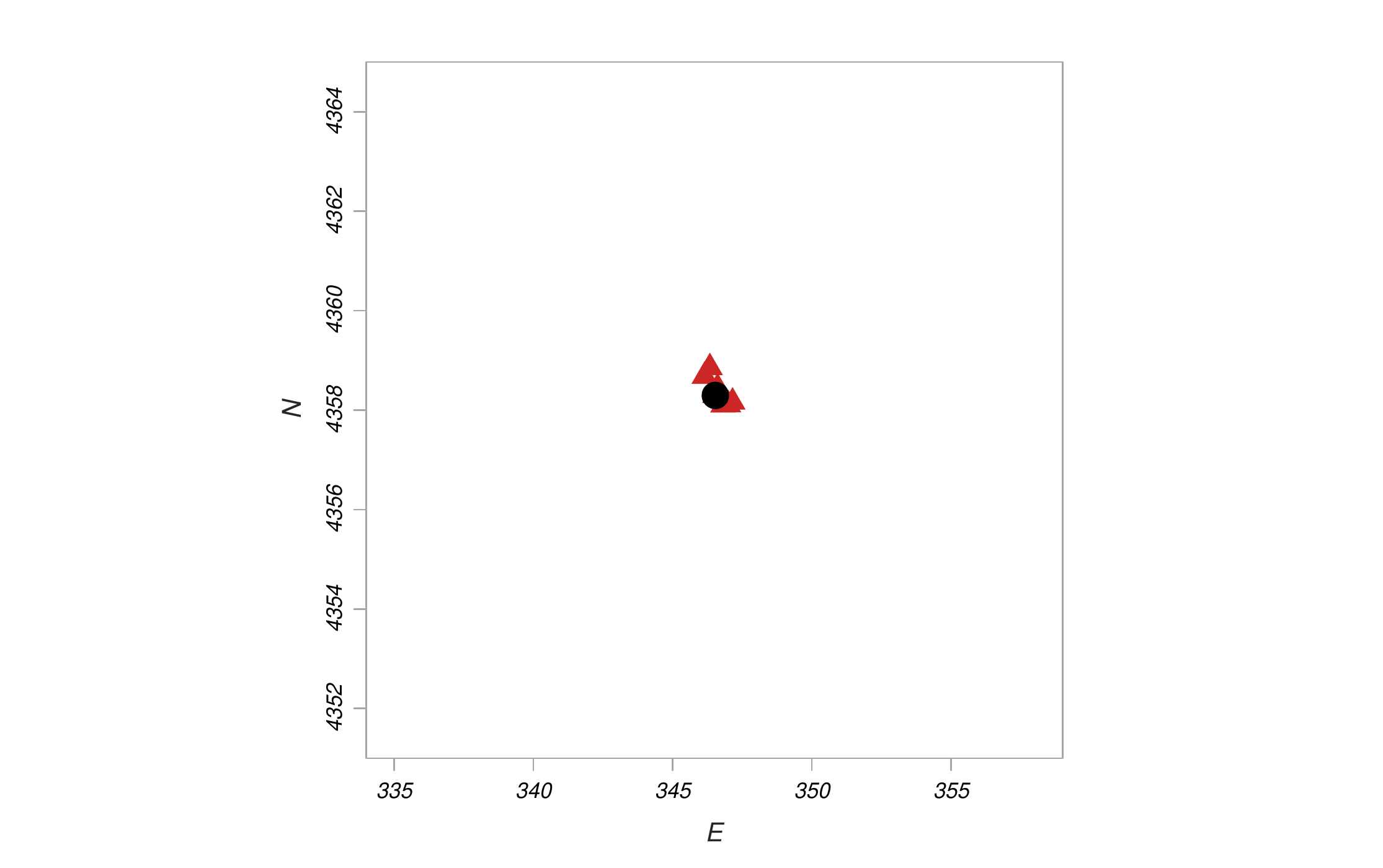}
                 \caption{\textit{Without a buffer zone.}}
                 \label{marr:1}
         \end{subfigure}~
    \begin{subfigure}[b]{0.49\textwidth}
                 \includegraphics[trim= 0mm 0mm 0mm 0mm,width=\textwidth]{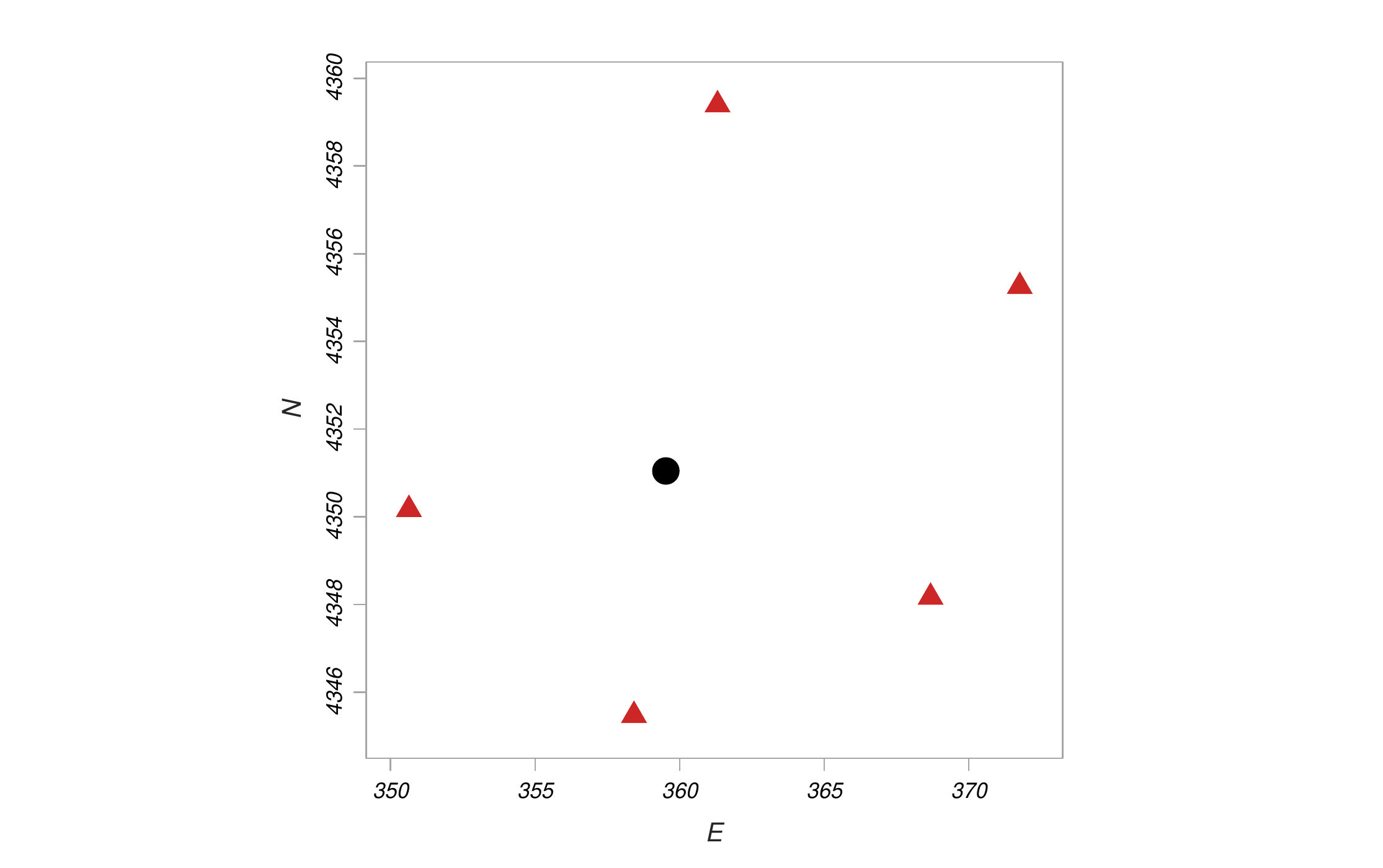}
                 \caption{\textit{With a buffer zone.}}
                 \label{marr:2}
         \end{subfigure}
\caption{\textit{Two basic subtypes of residents (red triangle indicates the crime site, the black circle denotes the anchor point).}}
      \label{marr1}
    \end{center}
  \end{figure}

The two basic subtypes of residents (Fig.~\ref{marr1}) are characterized by the existence of a buffer zone. 
In both cases the distribution of crime sites is depicted on an approximately same area $ 25\,\text{km}\times 15\,\text{km} $. However, 
in the first case  the offender commits crime in an immediate vicinity of the anchor point, having no buffer zone (see Fig. \ref{marr:1}).
In the second case, the offender commits crimes several kilometres from the anchor point, hence there is 
a buffer zone (see Fig. \ref{marr:2}). 

It was generally observed that if a resident has a small distance between crime sites (up to 2~km), the offender 
does not consider any buffer zone, and the behaviour is similar to the offender from Fig. \ref{marr:1}. Conversely, 
if the crime sites have larger distances and the sites are irregularly spaced,  
the offender's behaviour is similar to that of Fig. \ref{marr:2} and we have to take a buffer zone into account.

\begin{figure}[h]
    \begin{center}
    \begin{subfigure}[b]{0.49\textwidth}
                 \includegraphics[trim= 0mm 0mm 0mm 0mm,width=\textwidth]{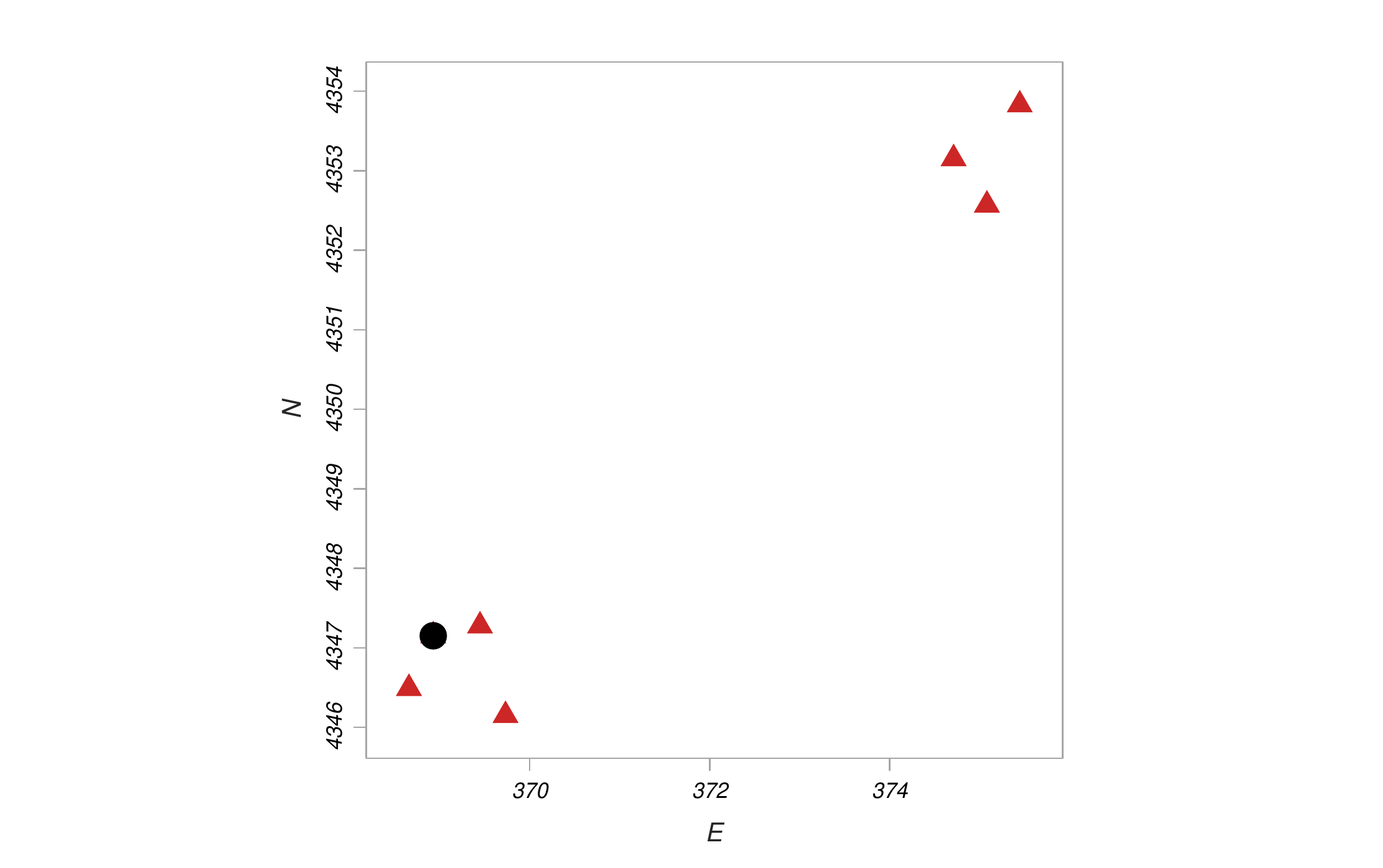}
                 \caption{\textit{Without a significant buffer zone.}}
                 \label{marr:3}
         \end{subfigure}~
    \begin{subfigure}[b]{0.49\textwidth}
                 \includegraphics[trim= 0mm 0mm 0mm 0mm,width=\textwidth]{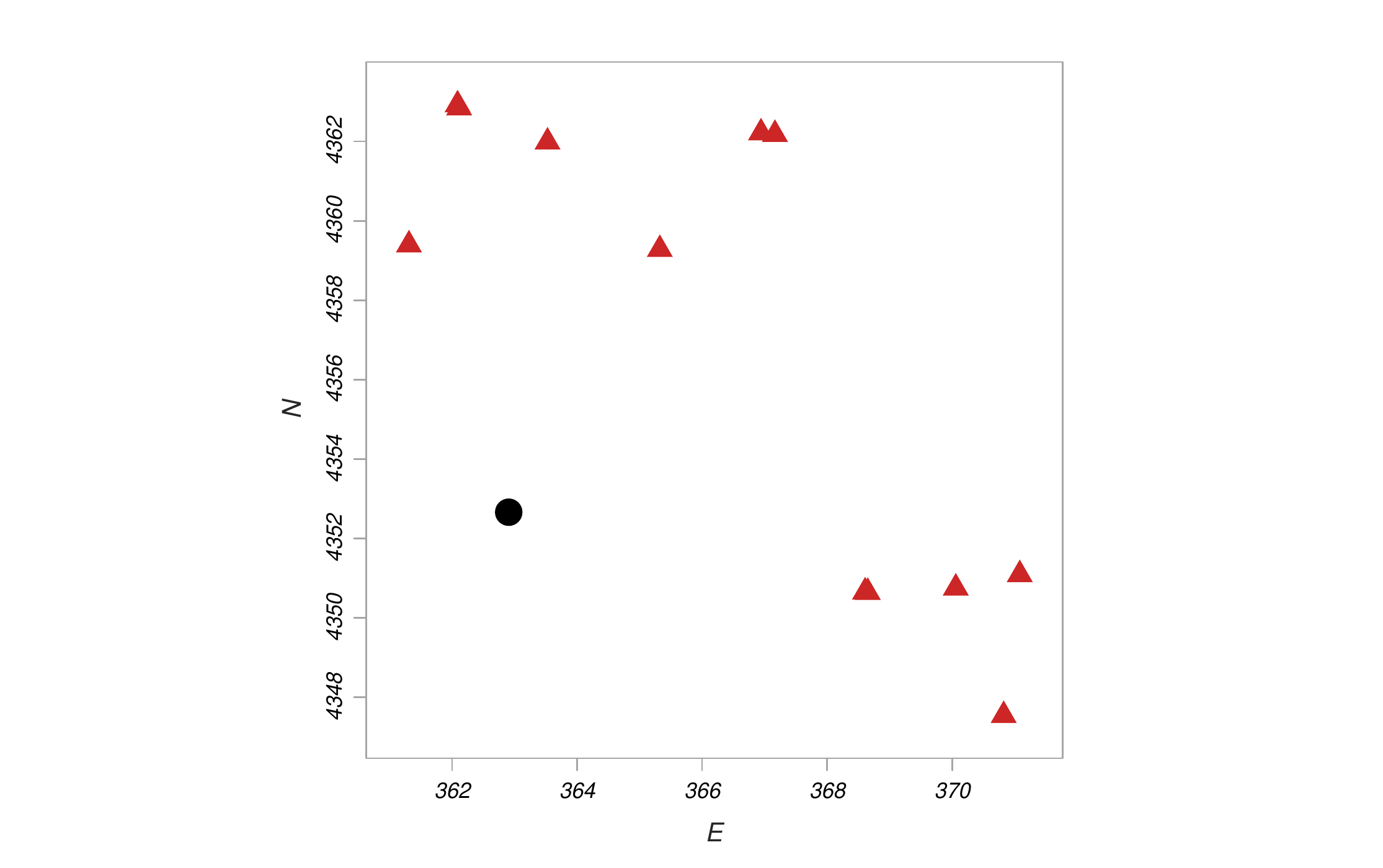}
                 \caption{\textit{With a buffer zone.}}
                 \label{marr:4}
         \end{subfigure}
\caption{\textit{Residents whose crime sites create clusters (the red triangle indicates the crime site, the black circle denotes the anchor point)).}}
      \label{marr2}
    \end{center}
  \end{figure}

We can also observe residents whose crime sites create clusters (see Fig. \ref{marr2}), where the distances  are small within some clusters,
but larger within others.

Some residents with crime sites in clusters have similar behavior as in Fig. \ref{marr:3}. They commit some crimes close to their anchor points, but other crimes in larger distances. Distant crime sites can, but do not have to, create clusters. On the other hand, some offenders  (see Fig.~\ref{marr:4}) create clusters although their anchor point does not occur in any of them.  Such criminals only prefer some locations. Therefore, if the crime sites of an offender create  clusters or a cluster with other isolated and distant crime sites, the offender can, but does not have to, create a buffer zone. In modelling we have to take this fact into account.

Undoubtedly, each criminal has an individuality with a unique behavior. Thus, we could find various specific features for each of them. However, the detailed examination of our data set shows that  each offender significantly tends to one of these subtypes. According to the space distribution of the criminal's crime sites we can decide which subtype is the most suitable for the investigated offender. However, in the case of clusters we are not able to decide on the existence of the criminal's buffer zone only on the basis of the space distribution of his crime sites.

We will apply various modifications of the normal distribution for residents modelling. Compared to the use of the exponential distribution, the results are not significantly different. We use the Euclidean metric for all models because we, thus, obtain an analytical expression.

\subsubsection{Residents without a buffer zone and without clusters (subtype M1)}

For offenders whose behavior is similar to Fig. \ref{marr:1}, we will use a model suggested in \citep{oleary1}. Using the Euclidean metric we obtain
\begin{equation}\label{E:marr_mod_1}
p\left( \mathbf{x}_{i}|\mathbf{z},\alpha\right)=\frac{1}{4\alpha^{2}}\cdot\exp\left( -\frac{\pi}{4\alpha^{2}}\,\left[\left(x_{i}^{\left(1 \right)}-z^{\left(1 \right) } \right)^{2} +\left(x_{i}^{\left(2 \right)}-z^{\left(2 \right) } \right) ^{2} \right]\right)  \,,
\end{equation}
which corresponds to two dimensional normal distribution with the  mean at the anchor point and standard deviation  $ \sigma=\sqrt{\frac{2}{\pi}}\,\alpha $. The reason for this choice of $ \sigma $ can be found in~\citep{oleary2}. The parameter $ \alpha $ denotes the average distance that the offender is willing to travel to commit a crime.

The most probable is committing a crime directly at the anchor point or its immediate vicinity, which is exactly the behavior
we expect for this type of offenders. Let us  remark that committing a crime at the anchor point may seem strange, but 
is not at all impossible. In fact, in our data set this situation occurs quite often, since the anchor point  is not 
necessarily the offender's place of residence, but his favourite bar, workplace, etc.

\subsubsection{Residents with a buffer zone and without clusters (subtype M2)}

For an offender as in Fig. \ref{marr:2}, the highest probability of the occurrence of crimes will not be at the anchor point, but some distance $ \alpha $ away 
from the anchor point, where $ \alpha $ denote the average distance to the crime site. This is captured by a probability distribution
of the form 
\begin{equation}\label{E:marr_mod_2}
p\left( \mathbf{x}_{i}|\mathbf{z},\alpha,\sigma\right)=\frac{1}{N\left(\alpha,\sigma \right) }\cdot\exp\left( -\frac{1}{2\sigma^{2}}\,\left[\sqrt{\left(x_{i}^{\left(1 \right)}-z^{\left(1 \right) } \right)^{2} +\left(x_{i}^{\left(2 \right)}-z^{\left(2 \right) } \right) ^{2}}-\alpha \right]^{2}\right)\,,
\end{equation}
where the choice of $ \sigma $ determines the decay of the probability as the distance varies from $ \alpha $.

In order to obtain a probability distribution, we must have
\begin{equation}
\iint\limits_{\mathbb{R}^{2}}\frac{1}{N\left(\alpha,\sigma \right) }\cdot\exp\left( -\frac{1}{2\sigma^{2}}\,\left[\sqrt{\left(x_{i}^{\left(1 \right)}-z^{\left(1 \right) } \right)^{2} +\left(x_{i}^{\left(2 \right)}-z^{\left(2 \right) } \right) ^{2}}-\alpha \right]^{2}\right)\,\dxij\dxid=1\,.
\end{equation} A simple calculation gives the normalizing factor
\begin{equation}\label{E:marr_normalizace}
N\left(\alpha,\sigma \right) =2\pi\sigma^{2}\cdot\exp\left(-\frac{\alpha^{2}}{2\sigma^{2}}\right) +2\pi\sqrt{2\pi}\alpha\sigma\left(1-\Phi\left(-\frac{\alpha}{\sigma} \right)  \right)\,, 
\end{equation}
where $ \Phi $  is the distribution function of the standard normal distribution.

\begin{figure}[h]
    \begin{center}
    \begin{subfigure}[b]{0.49\textwidth}
                 \includegraphics[trim= 65mm 20mm 65mm 25mm,width=\textwidth]{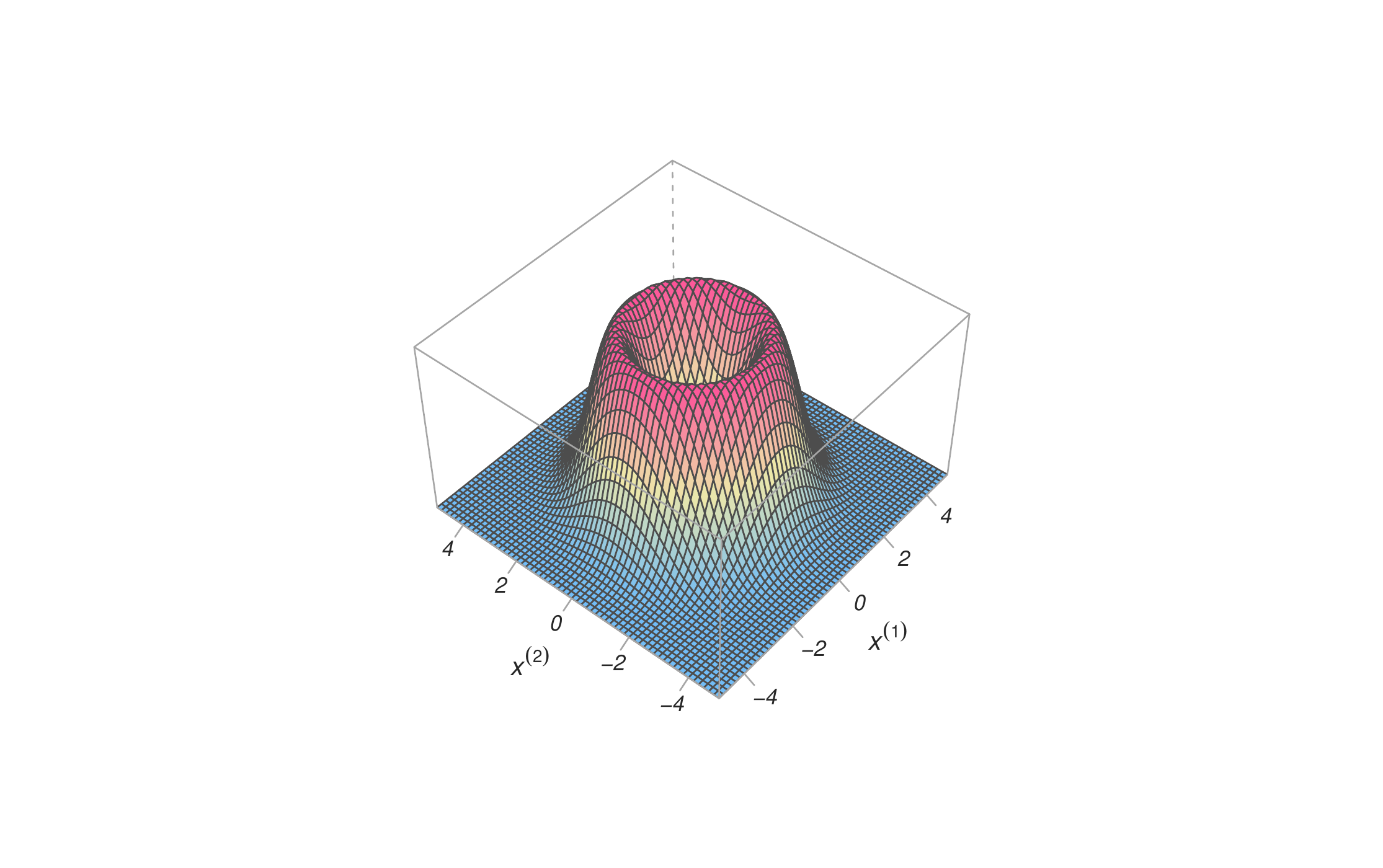}
                 \caption{\textit{Three-dimensional plot.}}
                 \label{mod_marr:3}
         \end{subfigure}~
    \begin{subfigure}[b]{0.49\textwidth}
                 \includegraphics[trim= 40mm 0mm 50mm 25mm,width=\textwidth]{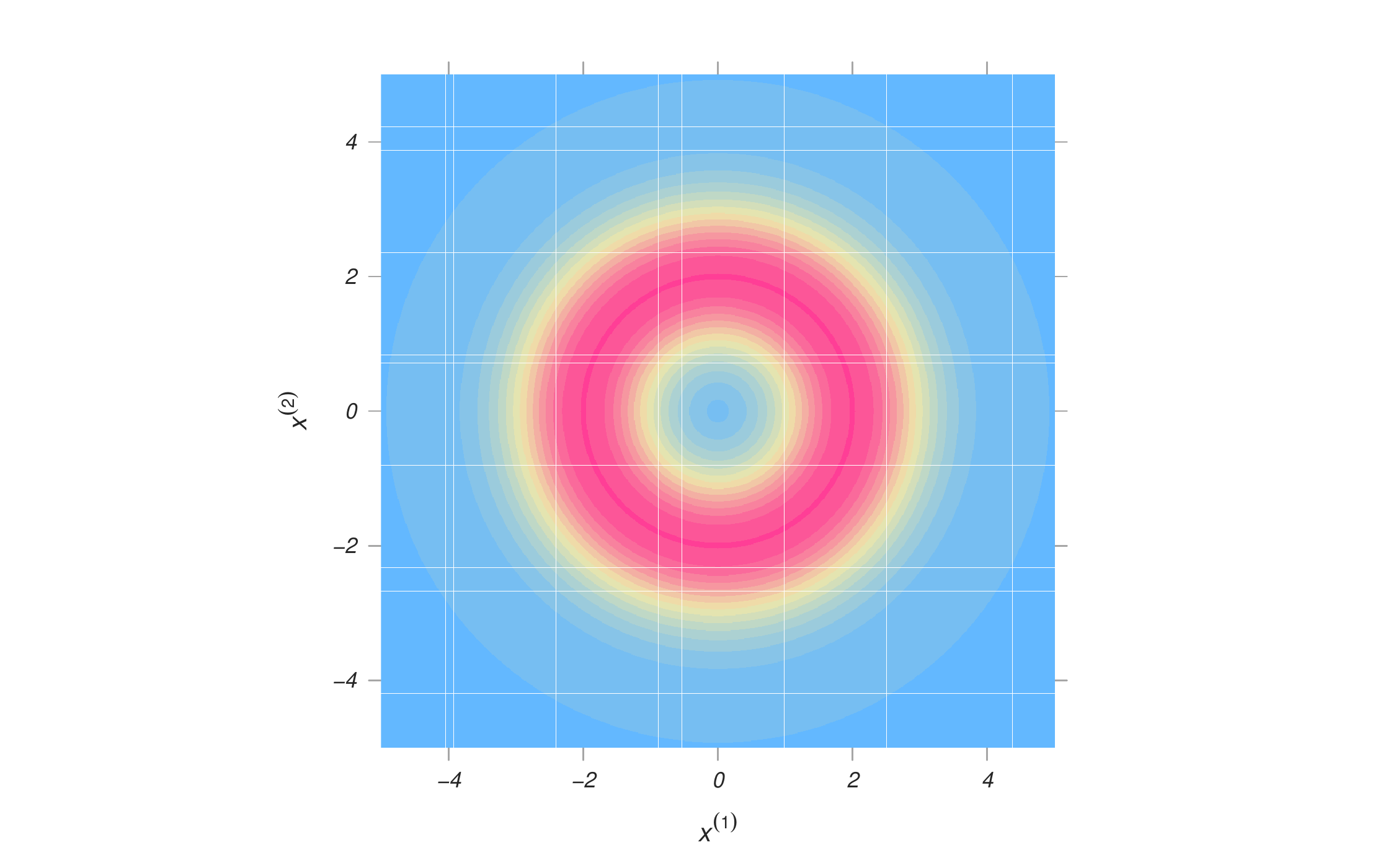}
                 \caption{\textit{Level plot.}}
                 \label{mod_marr:4}
         \end{subfigure}
\caption{\textit{Model for residents given by \eqref{E:marr_mod_2} with the anchor point $ \mathbf{z}=\left[0,0 \right]  $, $ \alpha=2 $ and~$ \sigma=0{,}8 $.}}
      \label{mod_marr2}
    \end{center}
  \end{figure}
  
Fig. \ref{mod_marr2} illustrates a model for this type of residents. The anchor point is $ \mathbf{z}=\left[0,0 \right]  $, $ \alpha=2 $ a~$ \sigma=0{,}8 $.  For different values of $ \alpha $ a~$ \sigma $,   
the probability of committing crime at the anchor point will be different, but nonzero, although for certain values 
very close to zero. 
The buffer zone does not have to be interpreted as an area where the criminal commits no crimes, but as 
an area around the anchor point, 
with a low probability of committing crimes. This interpretation corresponds better to the reality, where we 
can hardly claim with the certainty that there will be no crime in the buffer zone.

\bigskip

\noindent
\textbf{Remark.} \textit{As defined, these offenders are called residents because their anchor and criminal areas  overlap. However, they commit crimes at larger distances like non-residents (see Subsection \ref{dataCom}). If we choose an appropriate priors for  angle and distance, we can apply the model of non-residents (see \eqref{E:com_mod_1}) to this subtype of residents.}

\subsubsection{Residents with clusters (subtype M3)}
For offenders whose crime sites create clusters, as in Fig. ~\ref{marr2},  we are not able to decide 
whether their anchor point lies away from the clusters (they have a buffer zone), or whether it 
is in one of the clusters, and which one it is.

In this situation we will use multimodel inference \citep{burnham,oleary3} where we obtain 
the resulting estimate as a weighted average of the individual  estimates of anchor points. 

Let us assume that  we work with
$ R $ models for estimating the anchor point. We obtain  $ R $ posteriors, namely $ p_{i}\left(\mathbf{z}|\left\lbrace \mathbf{x}_{1},\mathbf{x}_{2},\ldots, \mathbf{x}_{n}\right\rbrace  \right)$ for~$ i=1,2,\ldots,R $. The resulting estimate for multimodel 
inference is then 

\begin{equation}\label{E:multimodel}
p\left(\mathbf{z}|\left\lbrace \mathbf{x}_{1},\mathbf{x}_{2},\ldots, \mathbf{x}_{n}\right\rbrace  \right)=\sum\limits_{i=1}^{R}\,w_{i}\cdot p_{i}\left(\mathbf{z}|\left\lbrace \mathbf{x}_{1},\mathbf{x}_{2},\ldots, \mathbf{x}_{n}\right\rbrace  \right)\,,
\end{equation}
where $ w_{i}\geqslant 0 $ denotes the weight corresponding to the $ i $-th estimate, and the weights satisfy $ \sum_{i=1}^{R}\,w_{i}=1 $. This  guarantees that 
the resulting estimate also provides a probability distribution.
For residents with $ R-1 $ clusters, we will construct one model for each cluster and another model for the situation that the offender has a buffer zone.

The choice of weights depends on the analyst.  If we have no reason to prefer one of the models, we simply set  $ w_{i}=\frac{1}{R} $ for all $ i $. We can also use the previous results and, with offenders of a similar type, we find out which of the models  is better describing
the offender's behaviour. According to the relativities we can then assign weights to individual models. It is also possible 
that some evidence during the investigation prefers one of the models. In this case we set the weights 
based on preferences of an experienced investigator.

\subsection{Models for non-residents}\label{dataCom}

\begin{figure}[h!]
    \begin{center}
\includegraphics [trim= 0mm 10mm 0mm  8mm, width=\textwidth]{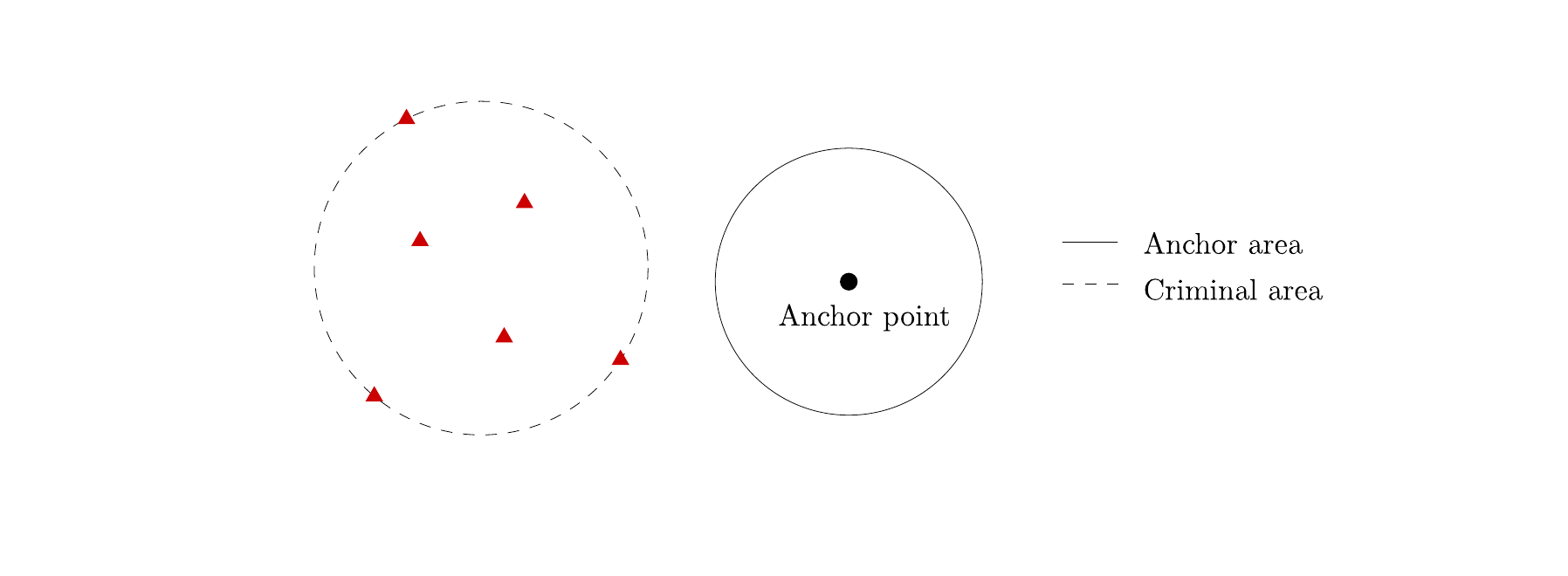}
\caption{\textit {Anchor and criminal areas of the non-resident offenders -- the dashed line circle denotes the criminal area, the solid line circle is the anchor area, red triangles indicate crime sites, the black circle is the anchor point of the offender.}}
      \label{commuters}
    \end{center}
  \end{figure}

A non-resident is an offender who commits crimes relatively far from his anchor point (see Fig. \ref{commuters}). 
 
The criminal and anchor areas essentially do not overlap in this case.  However, it does not mean that this area is unknown for the criminal. It is only significantly away from the territory in which he normally lives, works and acts as a "non-offender".

In our dataset, the non-residents typically commit crimes at a distance greater than 10~km, but it is often at least twice that distance. Another specific feature for this type of  offenders is that they prefer a certain angle (measured from the horizontal axis with the origin at the anchor point) for the choice of their crime location.

In our dataset, there is just a small number of non-residents and we did not observe any significantly different behavior among them. Therefore we do not divide this type of offenders into other subtypes as in the case of residents in the previous subsection.

In \citep{mohler2}, the authors use  a kinetic model that is derived on the basis of the following stochastic differential equation 

\begin{equation}\label{stoch_dif}
\dxt=\bm{\mu}\left( \mathbf{x}_t\right) +\sqrt{2D}\,\dwt\,,
\end{equation}
where $ \mathbf{x}\left(t \right)  $ denotes the position of the offender at time  $ t $, $ \mathbf{W}_t $ is a two-dimensional standard Wiener process, $ D $ denotes the diffusion parameter and ~$\bm{\mu}  $ is the drift. The drift term could be used to describe more complex criminal behavior. 

Since we can choose parameter values, the model is suitable for various types of  offenders. If we want to obtain the most realistic model which is applicable to  non-residents, we have to use numerical methods. However, the solution can be approximated by the product of a function of the distance and a function of the angle (again measured from the horizontal axis with the origin at the anchor point). This fact supports the idea that the criminal behavior could be generally modeled as the product of a suitable function which influences the most likely distance from the offender’s anchor point and another function that affects the probability of the angle preferred by the criminal. 

\begin{figure}[h]
    \begin{center}
    \begin{subfigure}[b]{0.49\textwidth}
                 \includegraphics[trim= 65mm 20mm 65mm 40mm,width=\textwidth]{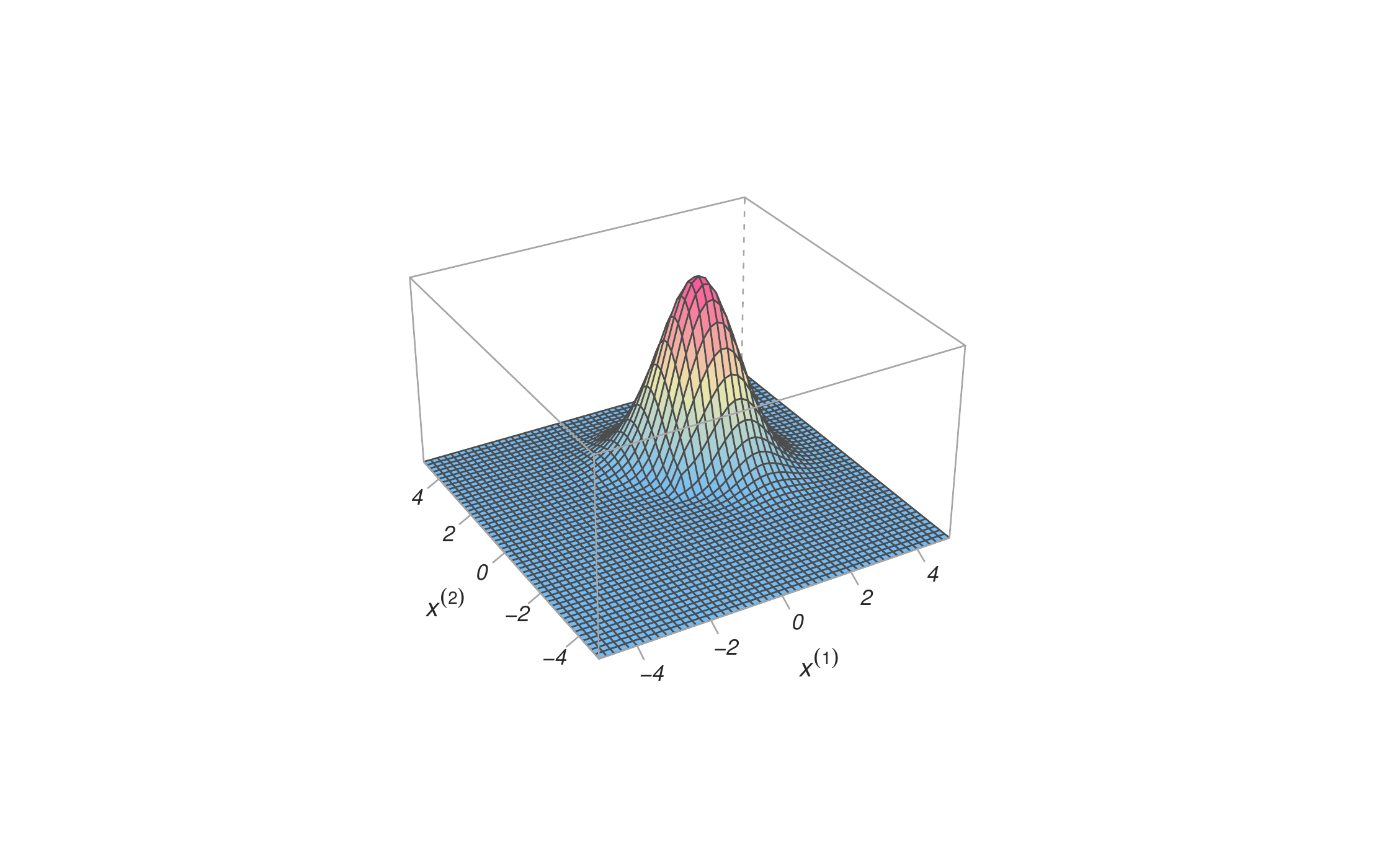}
                 \caption{\textit{Three-dimensional plot.}}
                 \label{mod_com:1}
         \end{subfigure}~
    \begin{subfigure}[b]{0.49\textwidth}
                 \includegraphics[trim= -10mm -10mm -10mm -10mm,width=\textwidth]{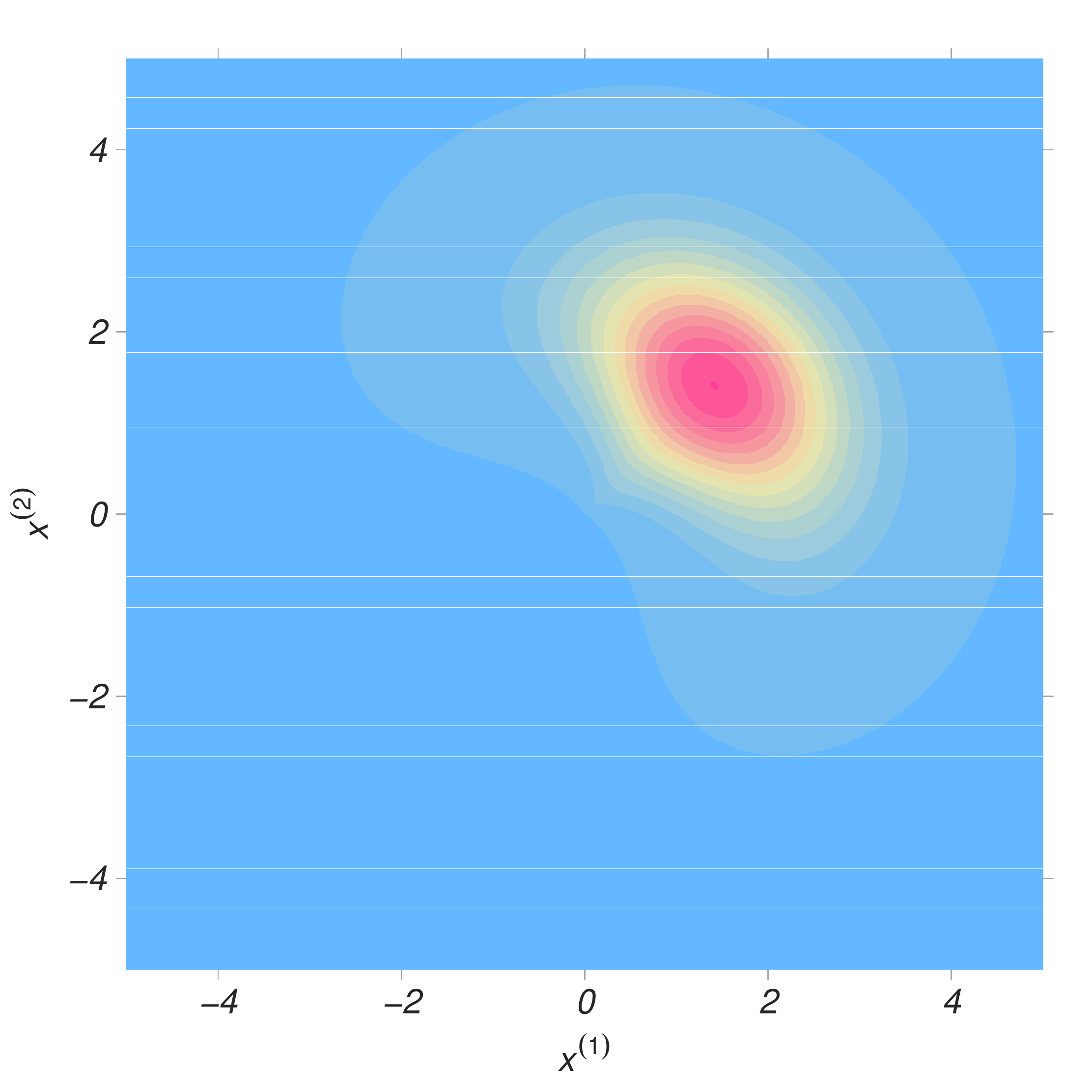}
                 \caption{\textit{Level plot.}}
                 \label{mod_com:2}
         \end{subfigure}
\caption{\textit{The function given by \eqref{E:com_mod_1} with the anchor point $ \mathbf{z}=\left[0,0 \right]  $, $ \alpha=2 $, $  \sigma_{1}=\frac{4}{5} $ ,$ \vartheta=\frac{\pi}{4} $ and~$ \sigma_{2}=\frac{\pi}{6} $.}}
      \label{O:mod_com1}
    \end{center}
  \end{figure}

Again, the model is based on the normal distribution and is given by
\begin{equation}\label{E:com_mod_1}
p\left( \mathbf{x}_{i}|\mathbf{z},\alpha,\vartheta,\sigma_{1},\sigma_{2}\right)=\frac{1}{N\left(\alpha,\vartheta,\sigma_{1},\sigma_{2} \right) }\cdot q_{1}\left(\mathbf{x}_{i}|\mathbf{z},\alpha,\sigma_{1} \right) \cdot
q_{2}\left(\mathbf{x}_{i}|\mathbf{z},\vartheta,\sigma_{2} \right)\,,
\end{equation}
where
\begin{equation}\label{E:q1}
q_{1}\left(\mathbf{x}_{i}|\mathbf{z},\alpha,\sigma_{1} \right)=\exp\left( -\frac{1}{2\sigma_{1}^{2}}\,\left[\sqrt{\left(x_{i}^{\left(1 \right)}-z^{\left(1 \right) } \right)^{2} +\left(x_{i}^{\left(2 \right)}-z^{\left(2 \right) } \right) ^{2}}-\alpha \right]^{2}\right)
\end{equation}
and
\begin{equation}\label{E:q2}
q_{2}\left(\mathbf{x}_{i}|\mathbf{z},\vartheta,\sigma_{2} \right)=\exp\left( -\frac{1}{2\sigma_{2}^{2}}\,\left[\mathrm{arg}\left(\left( x_{i}^{\left(1 \right)}-z^{\left(1 \right) }\right) +\mathrm{i}\left(  x_{i}^{\left(2 \right)}-z^{\left(2 \right) }\right) \right) -\vartheta \right]^{2}\right)\,,
\end{equation}
where $ \alpha $ is the average distance of the offenses,  $ \sigma_{1} $ denotes the standard deviation corresponding to the function $ q_{1} $, $ \vartheta $ is the average angle from the anchor point to the crime locations measured from the horizontal axis with the origin at the anchor point, and ~$ \sigma_{2} $  denotes the standard deviation corresponding to the function $ q_{2} $ (see Fig. \ref{O:mod_com1}).

The use of the function  $ q_{1} $ is analogous to the model  \eqref{E:marr_mod_2} for residents with a buffer zone, and therefore it can be interpreted in the same way. The function  $ q_{2} $ achieves the highest values at the angle $ \vartheta $ and the functional values around this angle decrease at a rate that depends on the choice of the value of $ \sigma_{2} $. This function itself cannot be normalized since the double integral over all its possible values is  infinite. However, the product of   $ q_{1} $ and $ q_{2} $ in \eqref{E:com_mod_1} can already be normalized.
 
Let us note that the normalization factor has the form 
\begin{equation}\label{Cnorm}
N\left(\alpha,\vartheta,\sigma_{1},\sigma_{2} \right)=N_{1}\left(\alpha,\sigma_{1} \right)\cdot N_{2}\left(\vartheta,\sigma_{2} \right)\,,
\end{equation}
where 
\begin{equation}\label{Cnorm1}
N_{1}\left(\alpha,\sigma_{1} \right)=\sigma_{1}^{2}\cdot\exp\left(-\frac{\alpha^{2}}{2\sigma_{1}^{2}}\right) +\sqrt{2\pi}\alpha\sigma_{1}\left(1-\Phi\left(-\frac{\alpha}{\sigma_{1}} \right)  \right)
\end{equation}
and
\begin{equation}\label{Cnorm2}
N_{2}\left(\vartheta,\sigma_{2} \right)=\sigma_{2}\sqrt{2\pi}\left(\Phi\left(\frac{2\pi-\vartheta}{\sigma_{2}} \right)-\Phi\left(-\frac{\vartheta}{\sigma_{2}} \right)  \right)\,,
\end{equation}
where $ \Phi $ represents the distribution function of the standard normal distribution. We obtain this value of the normalization factor from the requirement that the function $ p $ in \eqref{E:com_mod_1} has to be a probability density, hence
\begin{equation}
\iint\limits_{\mathbb{R}^{2}}\frac{1}{N\left(\alpha,\vartheta,\sigma_{1},\sigma_{2} \right) }\cdot q_{1}\left(\mathbf{x}_{i}|\mathbf{z},\alpha,\sigma_{1} \right) \cdot
q_{2}\left(\mathbf{x}_{i}|\mathbf{z},\vartheta,\sigma_{2} \right)\,\dxij\dxid=1\,.
\end{equation}

\bigskip

If we know in advance whether the offender is a resident or a non-resident, we choose an appropriate function proposed above to model his behavior. For residents, we have to decide on the offender’s subtype based on the distribution of the criminal’s crime locations.

If we are not able to determine in advance whether the offender is a resident or a non-resident, we can use multimodel inference, as discussed in Subsection \ref{PodsecMarr}.

\section{The choice of a prior distribution}\index{Funkce! priorní}\label{5}

When processing the dataset, we work with offenders of different types and subtypes. Depending on a specific offender, we need to select appropriate priors for the parameters. In this study, we used kernel smoothing and logspline density estimation  \citep{kooperberg} which allows to limit the range of values that the parameter can take. We always assume that all information about each offender contained in the data set  is known to us. We only exclude knowledge about the examined offender. Although all information in the dataset is known to us, in the following paragraphs we will use the word “known” for the data of all offenders except an investigated criminal. 

\begin{figure}[h]
    \begin{center}
    \begin{subfigure}[b]{0.46\textwidth}
                 \includegraphics[trim= 70mm 30mm 60mm  35mm,width=\textwidth]{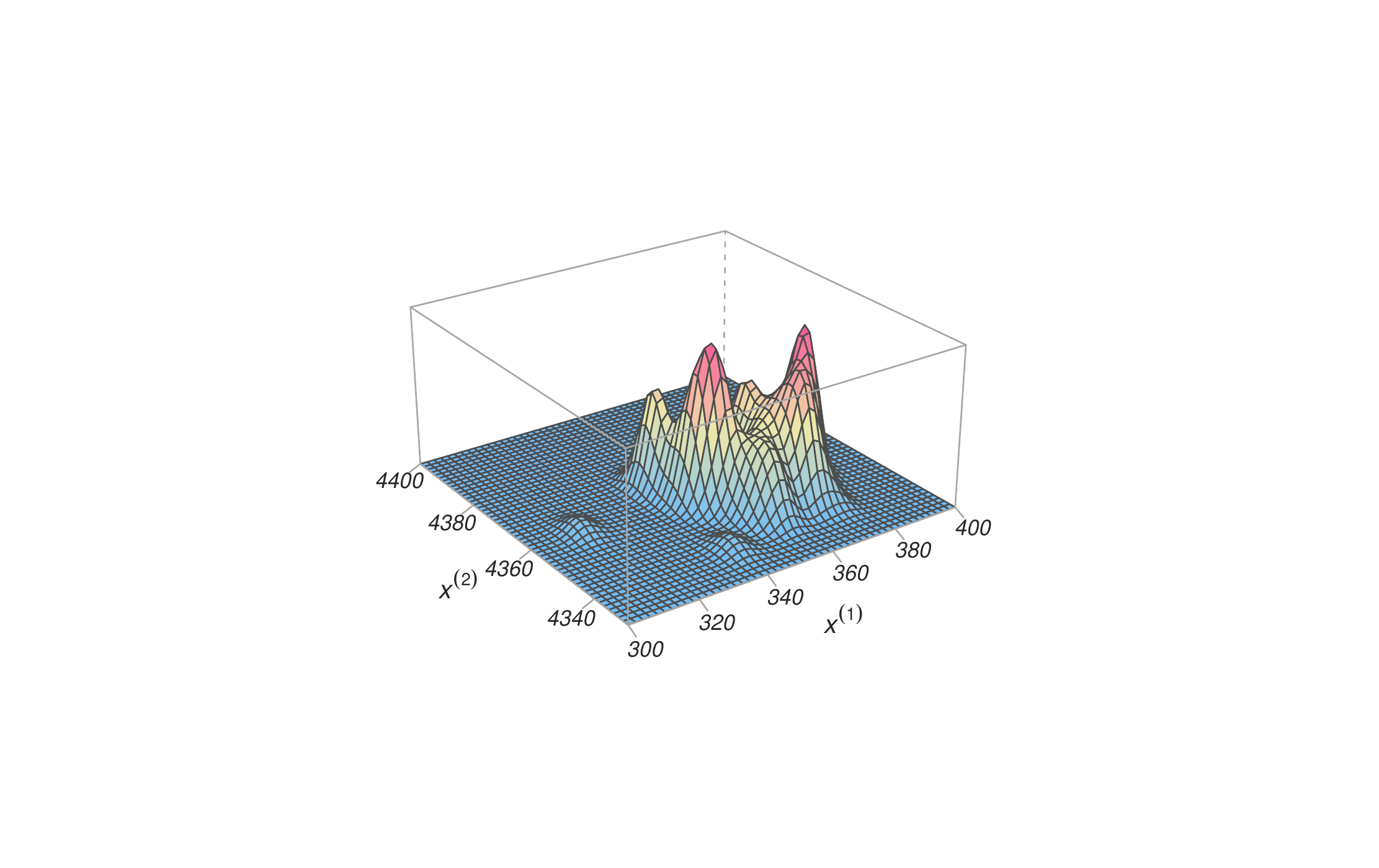}
                 \caption{\textit{Three-dimensional plot.}}
                 \label{O:PriorKot3D}
         \end{subfigure}~
    \begin{subfigure}[b]{0.52\textwidth}
                 \includegraphics[trim= 0mm -5mm 13mm  20mm,width=\textwidth]{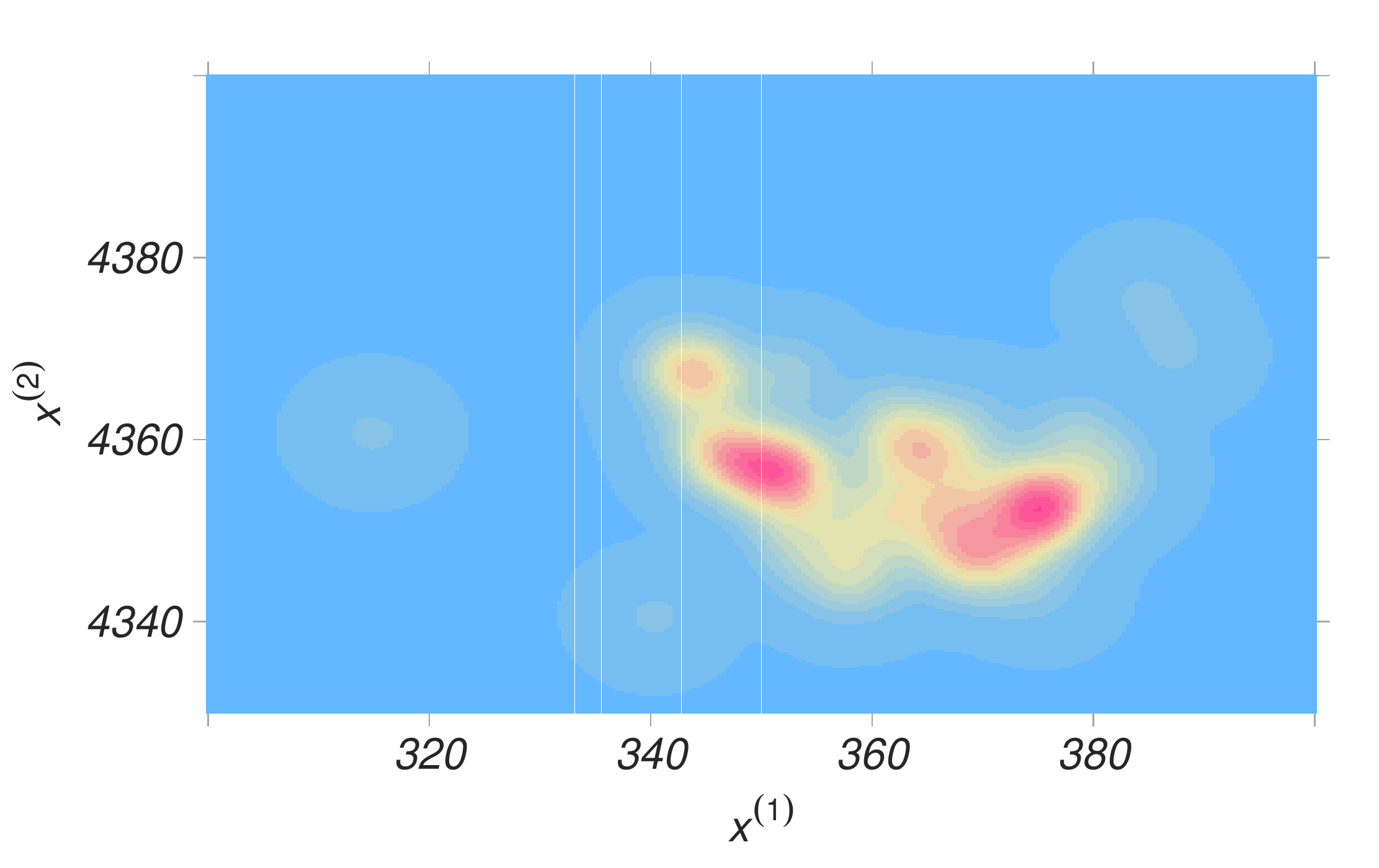}
                 \caption{\textit{Level plot.}}
                 \label{O:PriorKotVrst}
         \end{subfigure}
\caption{\textit{Graphical representation of the prior $ h\left(\mathbf{z} \right)  $ for the anchor point.}}
      \label{O:PriorKot}
    \end{center}
  \end{figure}

Fig. \ref{O:PriorKot} graphically  represents  our prior  for the anchor point. To obtain it, we used kernel smoothing, based on the anchor points of the known offenders.

\begin{figure}[h]
    \begin{center}
   \includegraphics[trim= 0mm 30mm 0mm  35mm,width=\textwidth]{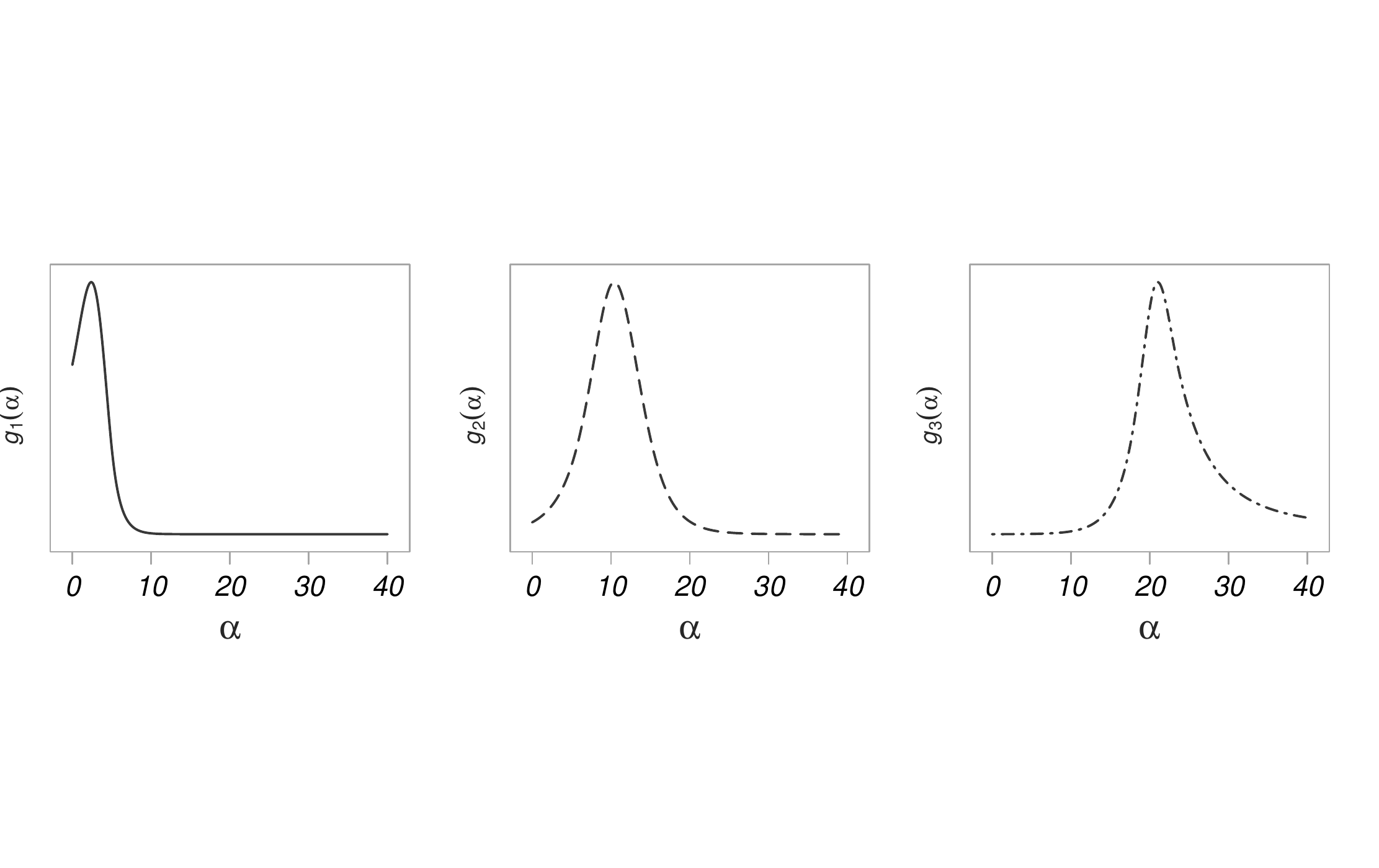}                
\caption{\textit{Graphical representation of the prior $ g_{1}\left(\alpha \right)  $ (solid line), $ g_{2}\left(\alpha \right) $ (dashed line) and~ $ g_{3}\left(\alpha \right) $ (dash-dotted line) for the average distance $ \alpha $ that the offender is willing to travel to commit a crime.}}
      \label{O:PriorAlfa}
    \end{center}
  \end{figure}

Next, we need to know the prior for the average distance  $ \alpha $ to the offence. Since the distance cannot take negative values we use logspline density estimation  with the lower limit equal to zero. In this case, however, we have to use solely the data corresponding to the particular types of offenders. This is because the distance is only one of the main factors which distinguish the types and subtypes of offenders one from another. The function  $ g_{1}\left(\alpha \right)  $ corresponds to the subtype M1, the $ g_{2}\left(\alpha \right)  $ to the subtype M2 and the function  $ g_{3}\left(\alpha \right)  $  is obtained from the known average distance for non-residents and  it will be used for this type of offenders. These functions are illustrated in Fig.  \ref{O:PriorAlfa}.

\begin{figure}[h]
    \begin{center}
   \includegraphics[trim= 0mm 25mm 0mm  30mm,width=\textwidth]{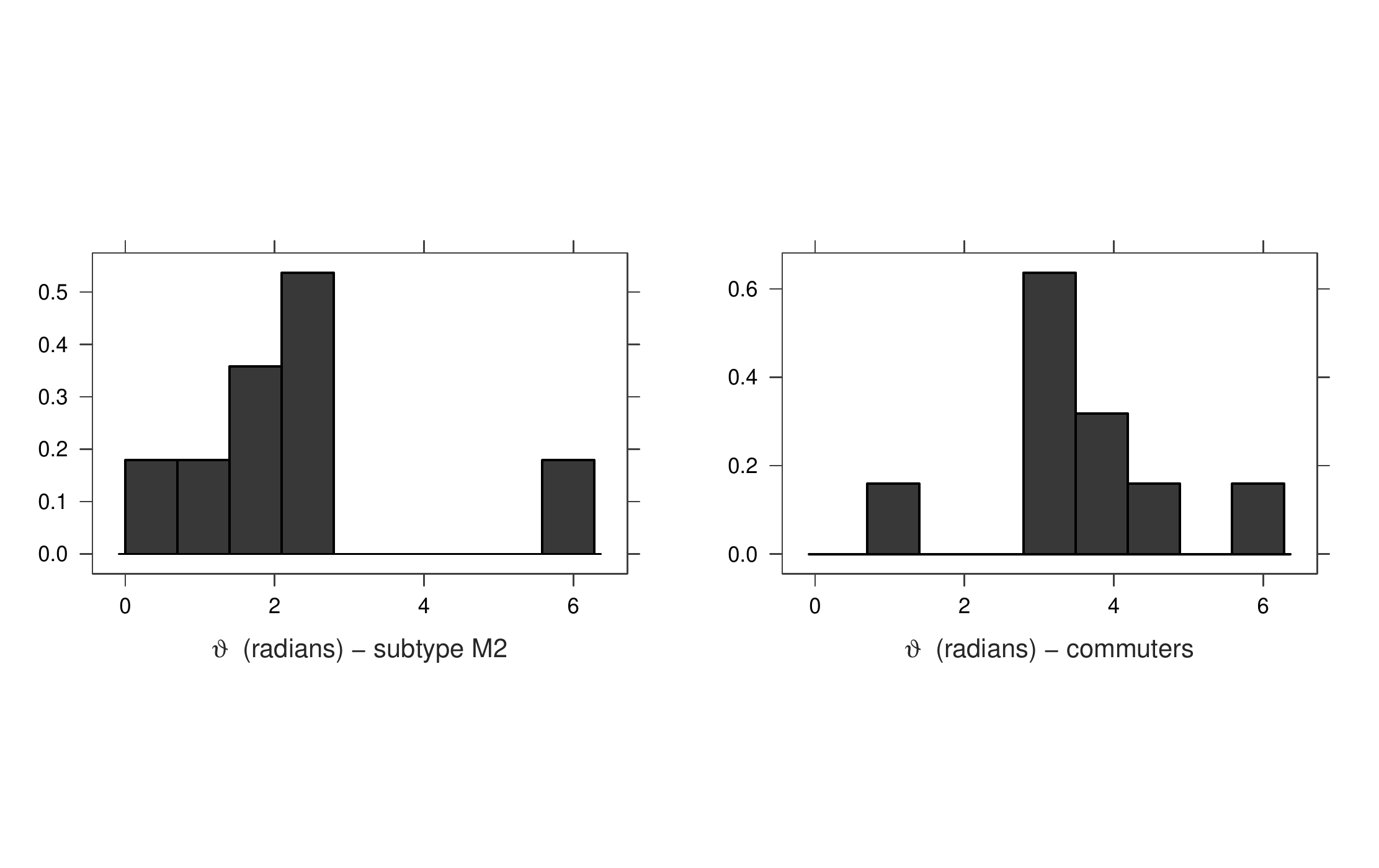}                
\caption{\textit{Histograms of the average angle $ \vartheta $ (measured from the horizontal axis with the origin at the anchor point) in which the offender commits crimes - for residents M2 and for non-residents.}}
      \label{O:PriorUhel}
    \end{center}
  \end{figure}

If we use the model given by \eqref{E:com_mod_1}, we have to know the prior distribution for the angle $ \vartheta $. In Fig. \ref{O:PriorUhel} we plot the estimated distribution of $ \vartheta $ for offenders of subtype M2 and for non-residents. We can see that offenders of subtype M2 prefer the angle between $ \frac{\pi}{2} $ and $ \pi $, non-residents favour more directions -- the south-west direction   (the angle between  ~$ \pi $ and $ \frac{3}{2}\pi $) and east direction  (the angle around   $ 0 $, or $ 2\pi $). The significant preference of some directions can be caused  by the existence of major transport networks for these angles. 

Similarly we obtain the prior distribution for other required parameters.

\section{Modelling and evaluation}\label{6}
In Section \ref{SecModData} we considered four types of offenders (residents M1, M2 and M3 and non-residents). For the modelling of each of these types, we can choose one of three models given by \eqref{E:marr_mod_1}, \eqref{E:marr_mod_2} and \eqref{E:com_mod_1}, or use multimodel inference, i. e. a combination of some of them.

Based on this, we will categorize the  modelling into the following cases:\label{metodyZnac}
\begin{enumerate}

\item We choose only residents from the dataset.
\begin{enumerate}
\renewcommand{\theenumi}{\arabic{enumi}}
\item\label{1a} 	We use the model \eqref{E:marr_mod_1} for the modelling of residents M1,  the model \eqref{E:marr_mod_2} for residents M2 and we use multimodel inference (a combination of  \eqref{E:marr_mod_1} and \eqref{E:marr_mod_2}) for residents M3.
\item\label{1b}
We use the model \eqref{E:marr_mod_1} for the modelling of residents M1,  the model \eqref{E:com_mod_1} for residents M2 and we use multimodel inference (a combination of  \eqref{E:marr_mod_1} and \eqref{E:com_mod_1}) for residents M3. 	

\end{enumerate}
\item 	We deal with all offenders without knowing in advance the type of the investigated offender. For each offender we admit the possibility that the offender is a resident of a certain type or a non-resident. Then we appropriately combine these two possibilities.
\begin{enumerate}

\item We use the model \eqref{E:marr_mod_1} for the modelling of residents M1,  the model  \eqref{E:marr_mod_2} for residents M2, multimodel inference (a combination of  \eqref{E:marr_mod_1} and  \eqref{E:marr_mod_2}) for residents M3 and the model \eqref{E:com_mod_1} for non-residents.
\begin{enumerate}

\item\label{2ai} 	Multimodelling weights are the same for both residents and non-residents.
\item\label{2aii} 	Multimodelling weights for residents and non-residents are derived by frequencies of these types in our dataset.
\end{enumerate}
\item We use the model \eqref{E:marr_mod_1} for the modelling of residents M1,  the model  \eqref{E:com_mod_1} for residents M2, multimodel inference (a combination of  \eqref{E:marr_mod_1} and  \eqref{E:com_mod_1}) for residents M3 and the model \eqref{E:com_mod_1} for non-residents.
\begin{enumerate}

\item\label{2bi} Multimodelling weights are the same  for both residents and non-residents.
\item\label{2bii} Multimodelling weights for residents and non-residents are derived by frequencies of these types in our dataset.
\end{enumerate} 
\end{enumerate}
\end{enumerate}

We compare these methods with Rossmo’s approach. Rossmo works with the hit score function given by  \eqref{SkorFunkce} and the distance decay function of the form
\begin{equation}
f\left( d\left( \mathbf{x}_{i} , \mathbf{y}  \right)\right) 
= \left\{ 
   \begin{array}{l l}
   
     \dfrac{k}{\left( d\left( \mathbf{x}_{i} , \mathbf{y}  \right) \right)^{h} } & \quad \text{for $ d\left( \mathbf{x}_{i} , \mathbf{y}  \right) > b$}\\

    \dfrac{kb^{g-h}}{\left(2b- d\left( \mathbf{x}_{i} , \mathbf{y}  \right) \right)^{g} } & \quad \text{for $ d\left( \mathbf{x}_{i} , \mathbf{y}  \right) \leqslant b$}\,,
     
   \end{array} \right. \,
\end{equation}
where $ b $ denotes the radius of the buffer zone, the distance $  d\left( \mathbf{x}_{i} , \mathbf{y}  \right) $ is calculated by the Manhattan metric and the exponents $ g $ and $ h $ are equal to  $ 1.2 $. These values of the parameters are recommended by Rossmo based on his research. We set the value of the parameter  $ b $ equal to one half of the average distance of the nearest neighbour between crimes in the given crime series. The choice of parameter  $ k $ is not important because the hit score function is the sum of the individual distance decay functions, the values of the hit score function are compared among themselves. The constant multiplies these values of the hit score function but does not change the ratios between them. Rossmo's  formula assumes that the offender’s anchor point is located close to his crime sites  (the fact how close depends on the optional parameter $ b $). This is the reason why this relationship is suitable especially for residents.

Let the investigated jurisdiction lie in the area of the rectangle with sides 100~km and 70~km, defined by the UTM coordinates 300~km west, 400~km east, 4330~km south and 4400~km north. The size of the jurisdiction was chosen to include all crimes and anchor points in the dataset $ \pm $ approximately 5~km. We divide this rectangle into a grid 70 $ \times $ 100, thus the dimensions of each cell are approximately 1,4~km $ \times $ 1,4~km.

We plot the crime locations for each investigated offender. Based on the space distribution, distances between crimes and occurrence or absence of clusters, we can determine the most appropriate criminal type, or subtype for the given offender. According to this type we choose a suitable model.

If we only deal with residents, in each cell of jurisdiction we evaluate the posterior for the considered methods described at the beginning of this section and for Rossmo’s approach. In  case  we examine all criminals without knowing the type (resident or non-resident) of the offender, we evaluate also the posterior  in each cell for the situation that the criminal is a  non-resident. Then we multiply the posterior for residents and the posterior for non-residents by the appropriate weights and after summing them we obtain an estimate of the anchor point  $ \mathbf{z} $, if the offender committed crimes in locations $ \mathbf{x}_{1},\mathbf{x}_{2},\ldots,\mathbf{x}_{n} $. Again, we apply this process to all methods described at the beginning of this section and compare them with Rossmo’s approach.

For each method, we order all cells based upon the value of the posterior, from the highest to lowest, thus from the cell that contains the anchor point with the highest probability to the cell that includes the anchor point with the lowest probability. The efficiency of the method depends on how many cells we have to examine until we find the anchor point of the investigated offender. If we divide this number of cells by the total number of cells in the given jurisdiction, we obtain the percentage of the area that we have to explore to find the offender’s anchor point. Thus, the method with the lowest percentage is the most effective for the particular series.

\begin{figure}[h]
    \begin{center}
    \begin{subfigure}[b]{0.49\textwidth}
                 \includegraphics[trim= 0mm 0mm 0mm  10mm,width=\textwidth]{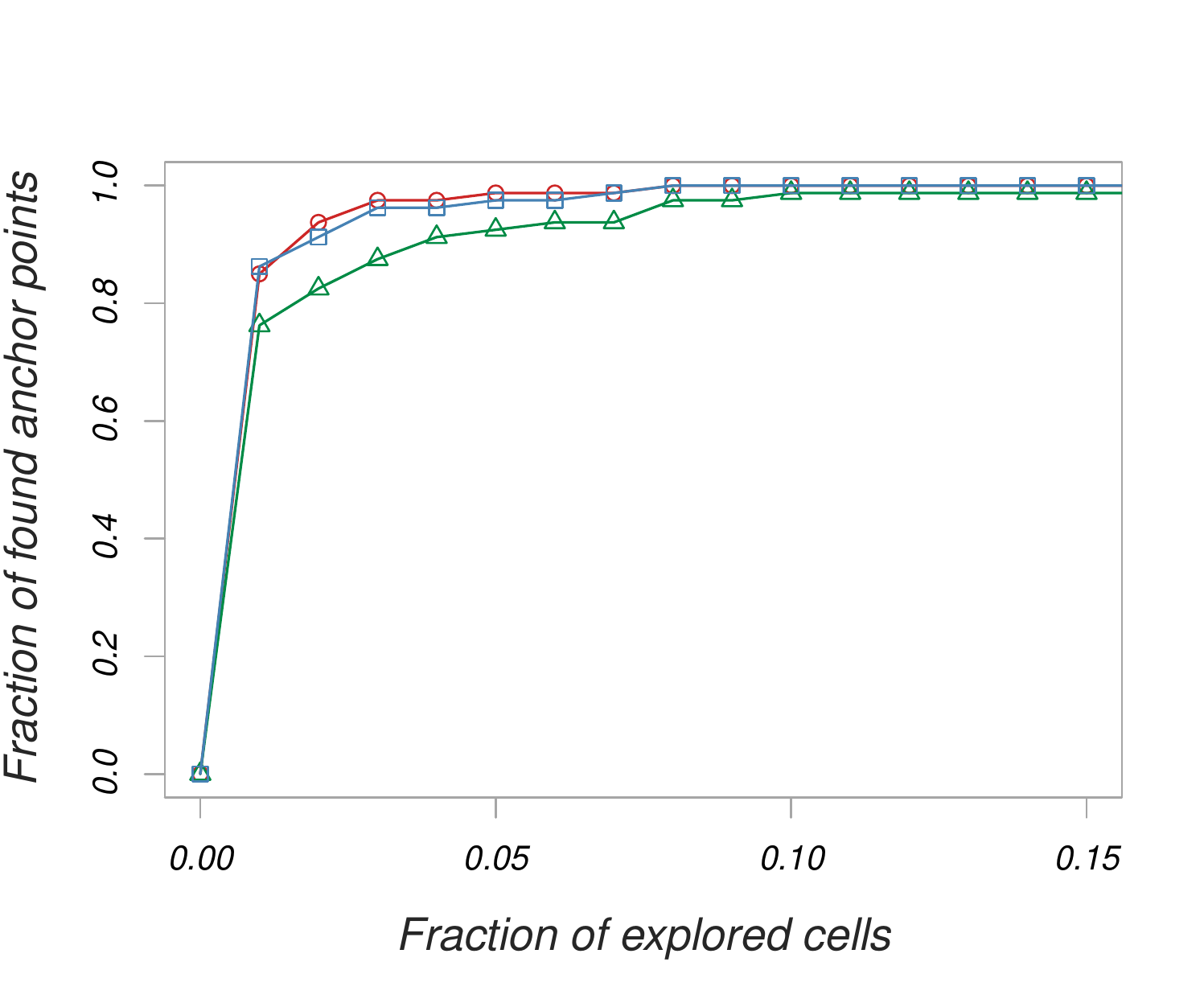}
                 \caption{\textit{Normal scale.}}
                 \label{O:Eval11}
         \end{subfigure}~
    \begin{subfigure}[b]{0.49\textwidth}
                 \includegraphics[trim= 0mm 0mm 0mm  10mm,width=\textwidth]{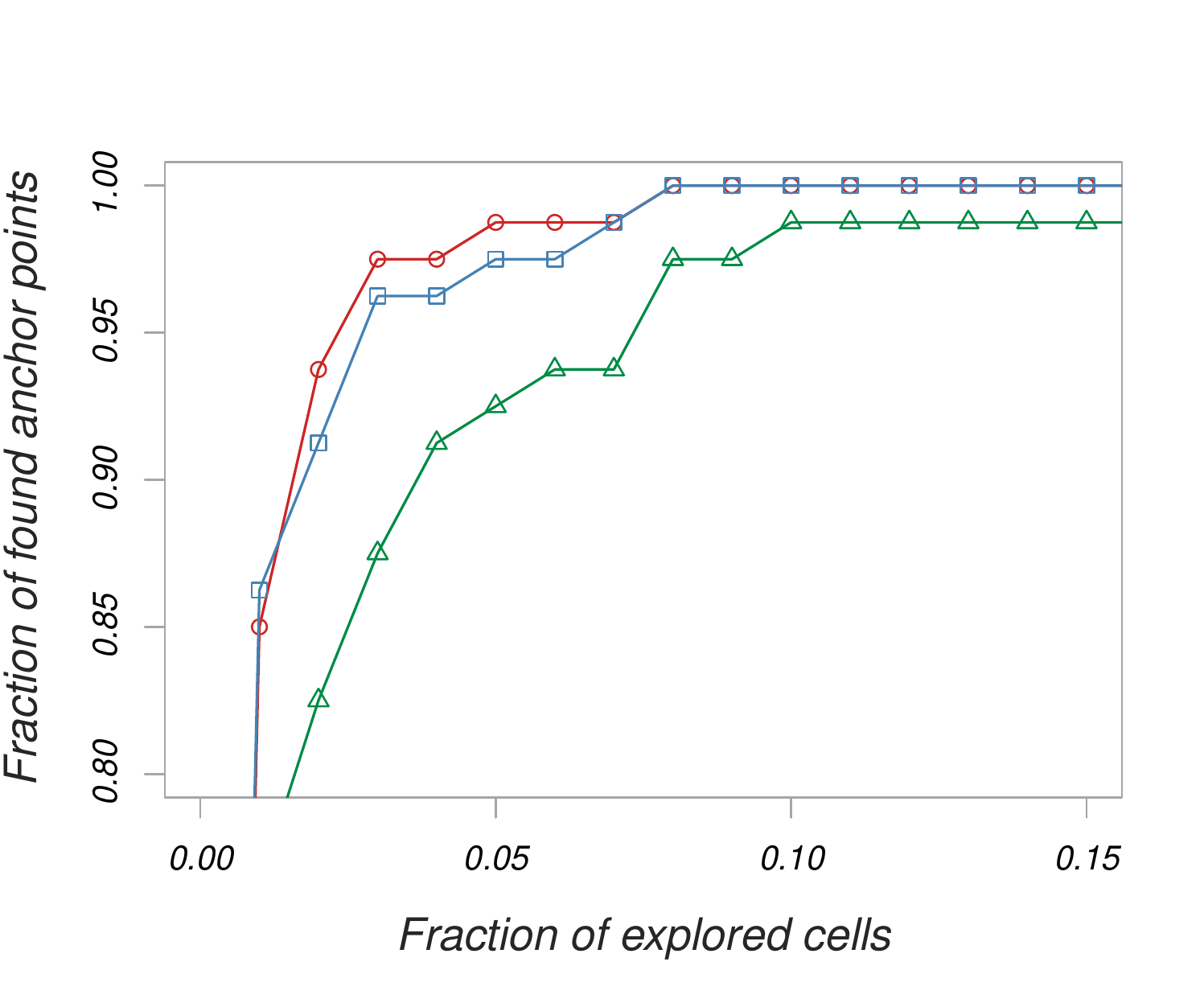}
                 \caption{\textit{Detail.}}
                 \label{O:Eval12}
         \end{subfigure}
               
\caption{\textit{The relationship between the proportion of the found anchor points and the proportion of the explored area, when we  deal with residents only; method \hyperref[1a]{1a} (red circles), method \hyperref[1b]{1b} (blue squares), Rossmo's approach (green triangles).}}
      \label{O:Eval1}
    \end{center}
  \end{figure}

Fig. \ref{O:Eval1} and Table \ref{tabulka1} show a comparison of methods \hyperref[1a]{1a}, \hyperref[1b]{1b} and Rossmo’s approach in terms of this  evaluation. We selected only the residents from the dataset; we did not consider the possibility that any of the criminals could be a non-resident. Although  Rossmo’s approach is appropriate just for residents, both methods \hyperref[1a]{1a} and~\hyperref[1b]{1b} indicate a  better, and very similar,  efficiency.  They had to  explore  only 8\% of all considered cells to find all anchor points, Rossmo's approach needed to explore 17\% of all cells to achieve the same.

\begin{table}[h]
\begin{center}

\noindent
\resizebox{\linewidth}{!}{%
\begin{tabular}{m{1cm}|>{\centering\arraybackslash}m{3cm}||c|c|c|c|c|c|c|c|c|c|c|c}
\hline 
& \textit{Percentage of explored cells} & \textit{1\% }& \textit{2\%} & \textit{3\%} & \textit{4\%} & \textit{5\%} & \textit{6\%} & \textit{7\%} & \textit{8\%} & \textit{9\%} & \textit{10\%} & \textit{16\%} & \textit{17\%} \\ 
\hline \hline &&&&&&&&&&&&&\\ [-0.5ex]
\multirow{3}{*}{\rotatebox[origin=c]{90}{\parbox[c]{18ex}{\centering \textit{Fraction of found anchor points}}}} & \textit{ Method \hyperref[1a]{1a}} & 0.8500 & \textbf{0.9375} & \textbf{0.9750} & \textbf{0.9750} & \textbf{0.9875} & \textbf{0.9875} & \textbf{0.9875} & \textbf{1} & \textbf{1} & \textbf{1} & \textbf{1} & \textbf{1} \\ [2ex]
\cline{2-14} &&&&&&&&&&&&&\\ [-0.5ex]
 & \textit{Method \hyperref[1b]{1b}} & \textbf{0.8625} & 0.9125 & 0.9625 & 0.9625 & 0.9750 & 0.9750 & \textbf{0.9875} & \textbf{1} & \textbf{1} & \textbf{1} & \textbf{1} & \textbf{1} \\ [2ex]
\cline{2-14}&&&&&&&&&&&&&\\ [-1.5ex]
 & \textit{Rossmo's approach} & 0.7625 & 0.8250 & 0.8750 & 0.9125 & 0.9250 & 0.9375 & 0.9375 & 0.9750 & 0.9750 & 0.9875 & 0.9875 & \textbf{1} \\ [2ex]
\hline 
\end{tabular} 
}

\end{center}
  \caption{\textit{The relationship between the proportion of the found anchor points and the percentage of the explored area, when we  deal with residents only (the method that found the most anchor points for a given percentage of explored cells is bold).}}
  \label{tabulka1}
\end{table}

\begin{figure}[h!]
    \begin{center}
    \begin{subfigure}[b]{0.49\textwidth}
                 \includegraphics[trim= 0mm 0mm 0mm  10mm,width=\textwidth]{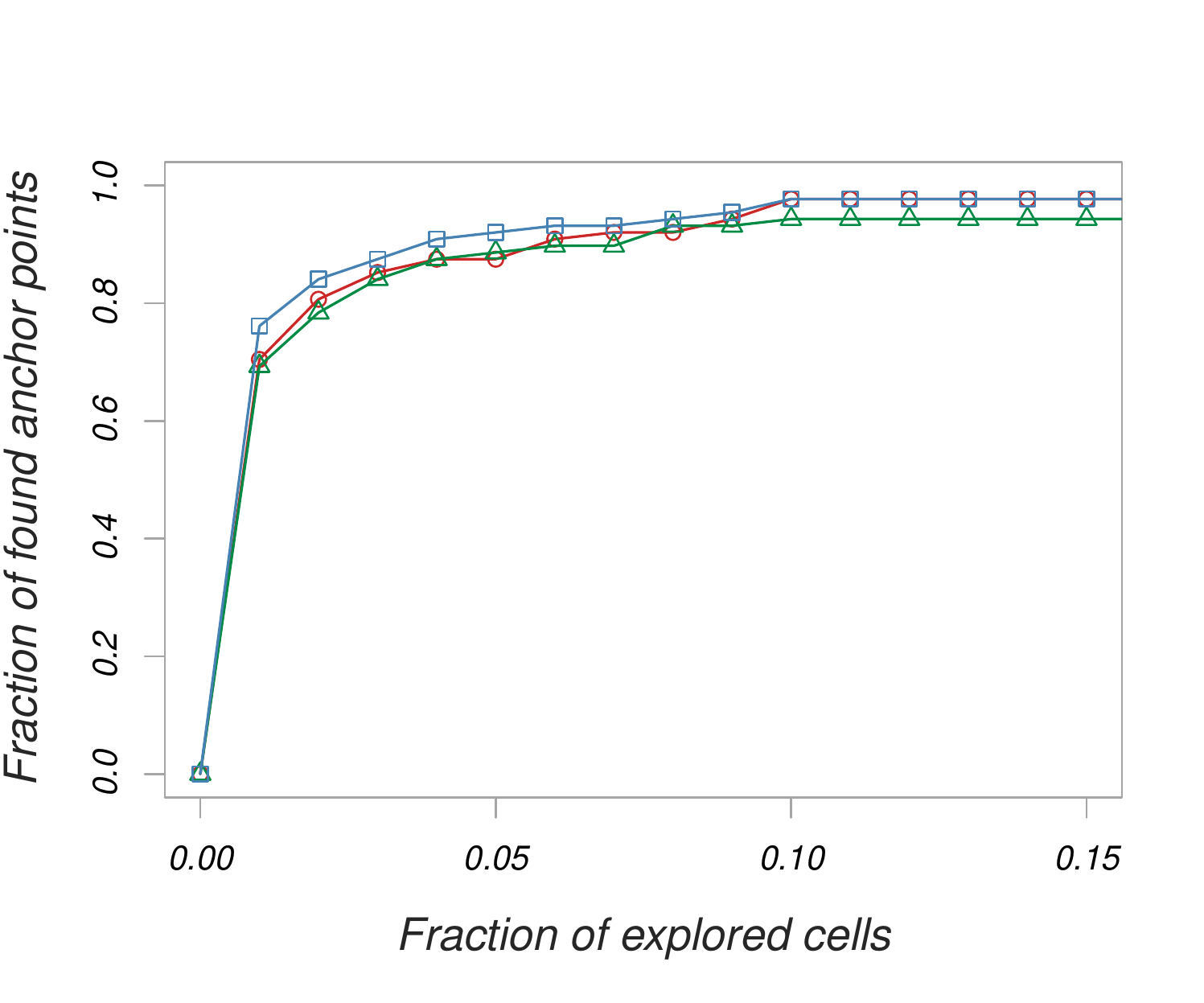}
                 \caption{\textit{The same weights for model of residents and model of non-residents.}}
                 \label{O:Eval2}
         \end{subfigure}~
    \begin{subfigure}[b]{0.49\textwidth}
                 \includegraphics[trim= 0mm 0mm 0mm  10mm,width=\textwidth]{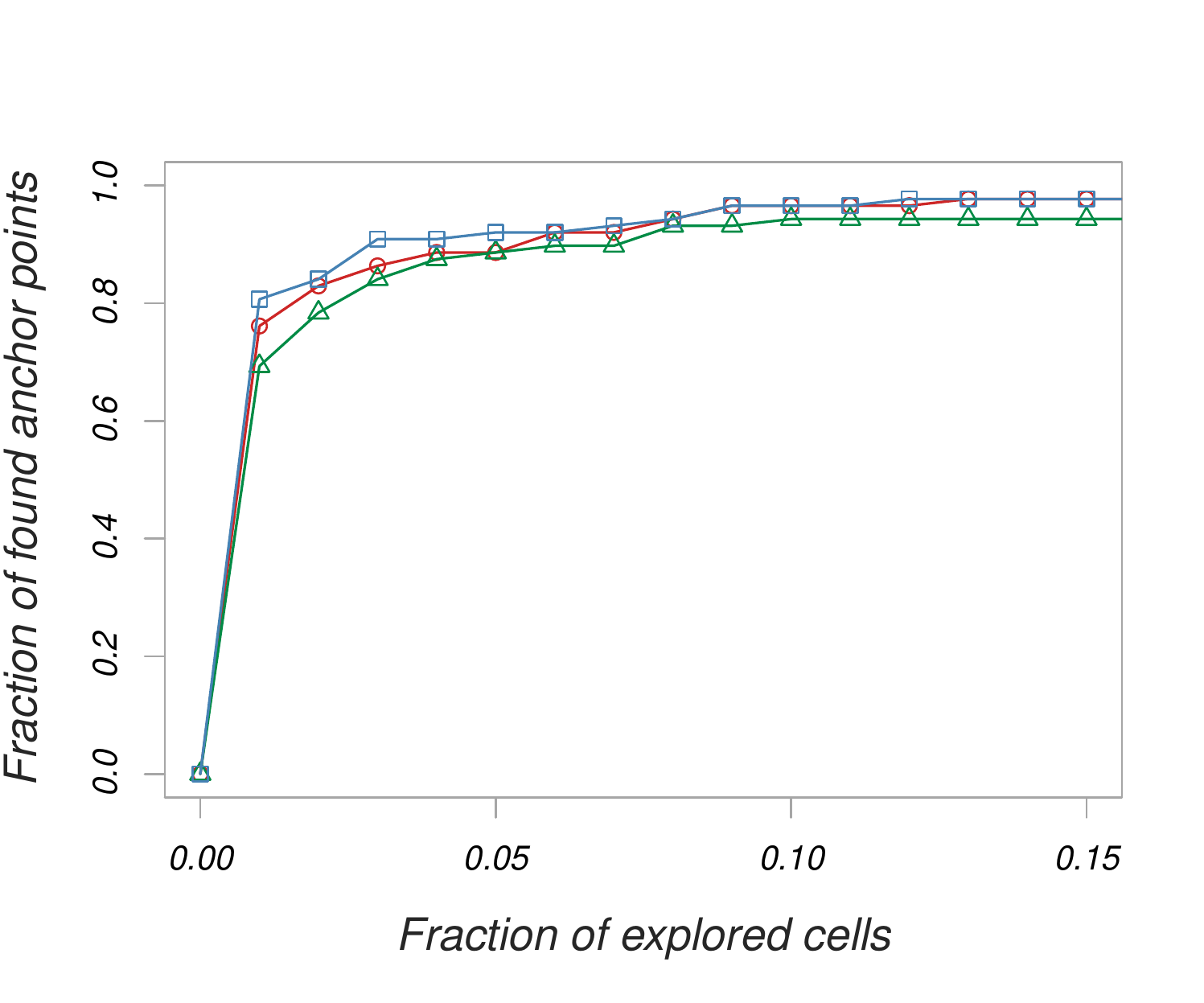}
                 \caption{\textit{The weights according to the frequencies of residents and non-residents in our data set.}}
                 \label{O:Eval3}
         \end{subfigure}
         
             \begin{subfigure}[b]{0.49\textwidth}
                 \includegraphics[trim= 0mm 0mm 0mm  5mm,width=\textwidth]{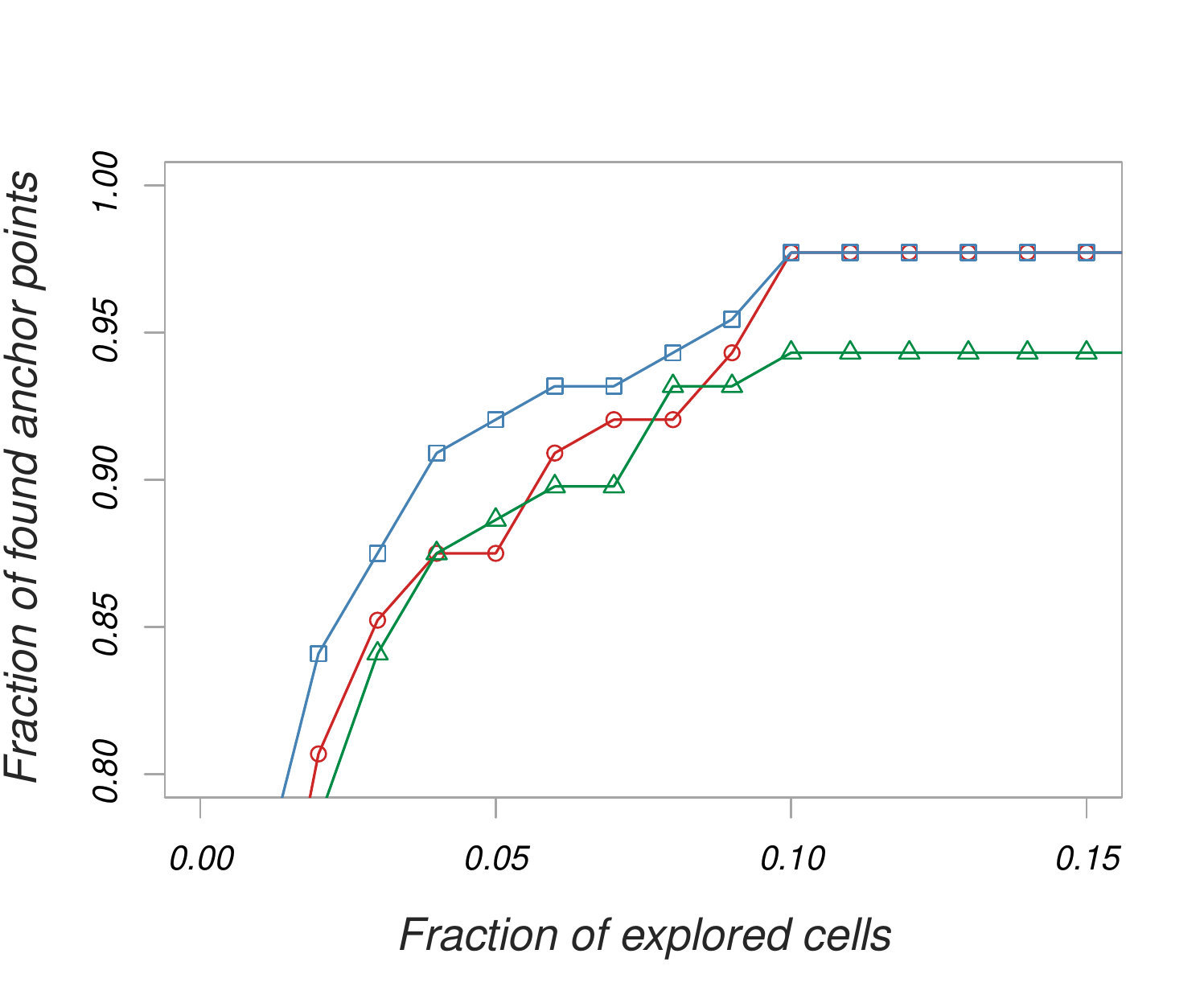}
                 \caption{\textit{ The same weights for model of residents and model of non-residents (detail).\\}}
                 \label{O:Eval4}
         \end{subfigure}~
    \begin{subfigure}[b]{0.49\textwidth}
                 \includegraphics[trim= 0mm 0mm 0mm  5mm,width=\textwidth]{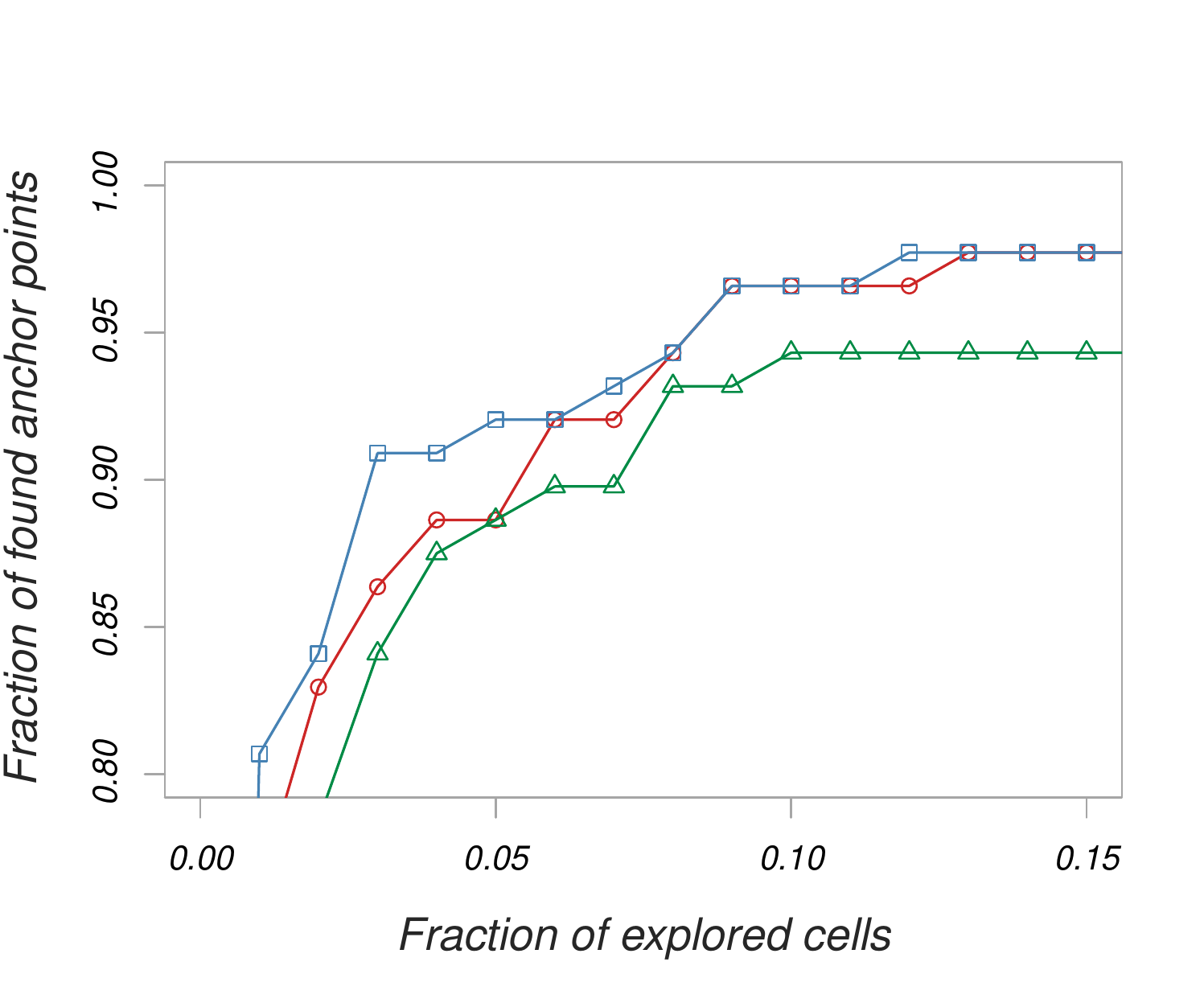}
                 \caption{\textit{The weights according to the frequencies of residents and  non-residents in our data set (detail).}}
                 \label{O:Eval5}
         \end{subfigure}
         
\caption{\textit{The relationship between the proportion of the found anchor points and the proportion of the explored area, when we deal with all offenders without knowing their types in advance. The parts (a) and (c): method \hyperref[2ai]{2ai} (red circles), method \hyperref[2bi]{2bi} (blue squares), Rossmo's approach (green triangles); The parts (b) and (d): method \hyperref[2aii]{2aii} (red circles), method \hyperref[2bii]{2bii} (blue squares), Rossmo's approach (green triangles).} }
      \label{O:Eval23}
    \end{center}
  \end{figure}

Fig. \ref{O:Eval23} and Table \ref{tabulka2} give a  comparison of methods  \hyperref[2ai]{2ai} and \hyperref[2bi]{2bi} and Rossmo’s approach (Fig.~\ref{O:Eval2} and Fig.~\ref{O:Eval4}) in terms of the described means of evaluation. It also  compares methods \hyperref[2aii]{2aii} and \hyperref[2bii]{2bii} and Rossmo's approach  (Fig.~\ref{O:Eval3} and  Fig.~\ref{O:Eval5}). Now we consider all offenders without knowing the type of each criminal. We can see that Rossmo's approach almost always exhibits the lowest efficiency in both cases.  Our four considered methods found all anchor points after exploring 37\% of all cells, Rossmo's approach needed to explore 39\% of all cells to find all anchor points. Moreover, Table \ref{tabulka2} shows that method \hyperref[2bii]{2bii} finds 10\% more anchor points than Rossmo's approach just after exploring 1\% of all considered cells.  However, the difference between the efficiencies  of the methods is not as remarkable as in the situation considered above when we dealt only with residents. Due to the inclusion of all criminals to the analysis, the estimation of the anchor point for methods \hyperref[2ai]{2ai}, \hyperref[2bi]{2bi} and  \hyperref[2aii]{2aii}, \hyperref[2bii]{2bii} worsened for residents since we had to admit the possibility that the investigated offender is a  non-resident. These methods achieve better results for non-residents than Rossmo’s approach. However,  in our datasets, the non-residents constitute only $ \frac{1}{11} $ of all offenders (it corresponds to the choice of multimodel weights for methods \hyperref[2aii]{2aii} and \hyperref[2bii]{2bii}).

\begin{table}[h]
\begin{center}

\noindent
\resizebox{\linewidth}{!}{%
\begin{tabular}{m{0.5cm}|>{\centering\arraybackslash}m{3cm}||c|c|c|c|c|c|c|c|c|c}
\hline 
& \textit{Percentage of explored cells} & \textit{1\% }& \textit{5\%} & \textit{10\%} & \textit{15\%} & \textit{20\%} & \textit{25\%} & \textit{36\%} & \textit{37\%} & \textit{38\%} & \textit{39\%}  \\ 
\hline \hline &&&&&&&&&&&\\ [-0.5ex]
\multirow{6}{*}{\rotatebox[origin=c]{90}{\parbox[c]{30ex}{\centering \textit{Fraction of found anchor points}}}} & \textit{ Method \hyperref[2ai]{2ai}} & 0.7045 & 0.8750 &\textbf{ 0.9773} & \textbf{0.9773 }& \textbf{0.9773} & \textbf{0.9886} & \textbf{0.9886} & \textbf{1}  & \textbf{1} & \textbf{1} \\ [2ex]
\cline{2-12} &&&&&&&&&&&\\ [-0.5ex]
 & \textit{Method \hyperref[2bi]{2bi}} & 0.7614 & \textbf{0.9205} & \textbf{0.9773} & \textbf{0.9773} & \textbf{0.9773} &\textbf{ 0.9886} & \textbf{0.9886} & \textbf{1} & \textbf{1} & \textbf{1}  \\ [2ex]
\cline{2-12}&&&&&&&&&&&\\ [-1.5ex]
& \textit{ Method \hyperref[2aii]{2aii}} & 0.7614 & 0.8864 & 0.9660 & \textbf{0.9773} & \textbf{0.9773} & \textbf{0.9886 }& \textbf{0.9886} & \textbf{1}  & \textbf{1} & \textbf{1} \\ [2ex]
\cline{2-12} &&&&&&&&&&&\\ [-0.5ex]
 & \textit{Method \hyperref[2bii]{2bii}} & \textbf{0.8068} & \textbf{0.9205} & 0.9660 & \textbf{0.9773} & \textbf{0.9773} & \textbf{0.9886} & \textbf{0.9886 }& \textbf{1} & \textbf{1} & \textbf{1}  \\ [2ex]
\cline{2-12}&&&&&&&&&&&\\ [-1.5ex]
 & \textit{Rossmo's approach} & 0.7011 & 0.8966 & 0.9540& 0.9540 & 0.9770 & 0.9885 & 0.9885 & 0.9885 & 0.9885 & \textbf{1}  \\ [2ex]
\hline 
\end{tabular} 
}

\end{center}
  \caption{\textit{The relationship between the proportion of the found anchor points and the percentage of the explored area when we deal with all offenders without knowing their types in advance (the method that found the most anchor points for a given percentage of explored cells is bold).}}
  \label{tabulka2}
\end{table}

 Nevertheless, we can say that methods \hyperref[2ai]{2ai}, \hyperref[2bi]{2bi} and  \hyperref[2aii]{2aii}, \hyperref[2bii]{2bii} exhibit better results than Rossmo’s approach (although not significant due to the previous case when we dealt with residents only). We can assume even a higher efficiency for datasets with a greater proportion of non-residents. We can see that, for our dataset, the best choice for residents with a buffer zone is the model of the form \eqref{E:com_mod_1} in  both cases. The option of multimodel weights distinguishes the considered method in the expected way. If we assign the same weights to  both types of criminals  (Fig.~\ref{O:Eval2}, respectively Fig. \ref{O:Eval4}), non-residents are caught earlier. On the contrary, the choice of weights based on frequencies of the types in the dataset  (Fig.~\ref{O:Eval3}, respectively Fig. \ref{O:Eval5}), when the weight for the model of residents corresponds to the value of  $ w_{1}=\frac{10}{11} $ and for the model of non-residents takes the value of  $ w_{2}=\frac{1}{11} $, leads to earlier capturing of residents at the expense of  finding  non-residents later. Overall, Fig. \ref{O:Eval23}  shows that the use of the prior knowledge about the structure of the dataset for determining the weights results in  capturing  a larger part of the offenders faster.
 
\begin{figure}[t]
    \begin{center}
    \begin{subfigure}[b]{0.48\textwidth}
                 \includegraphics[trim= 0mm 0mm 0mm  0mm,width=\textwidth]{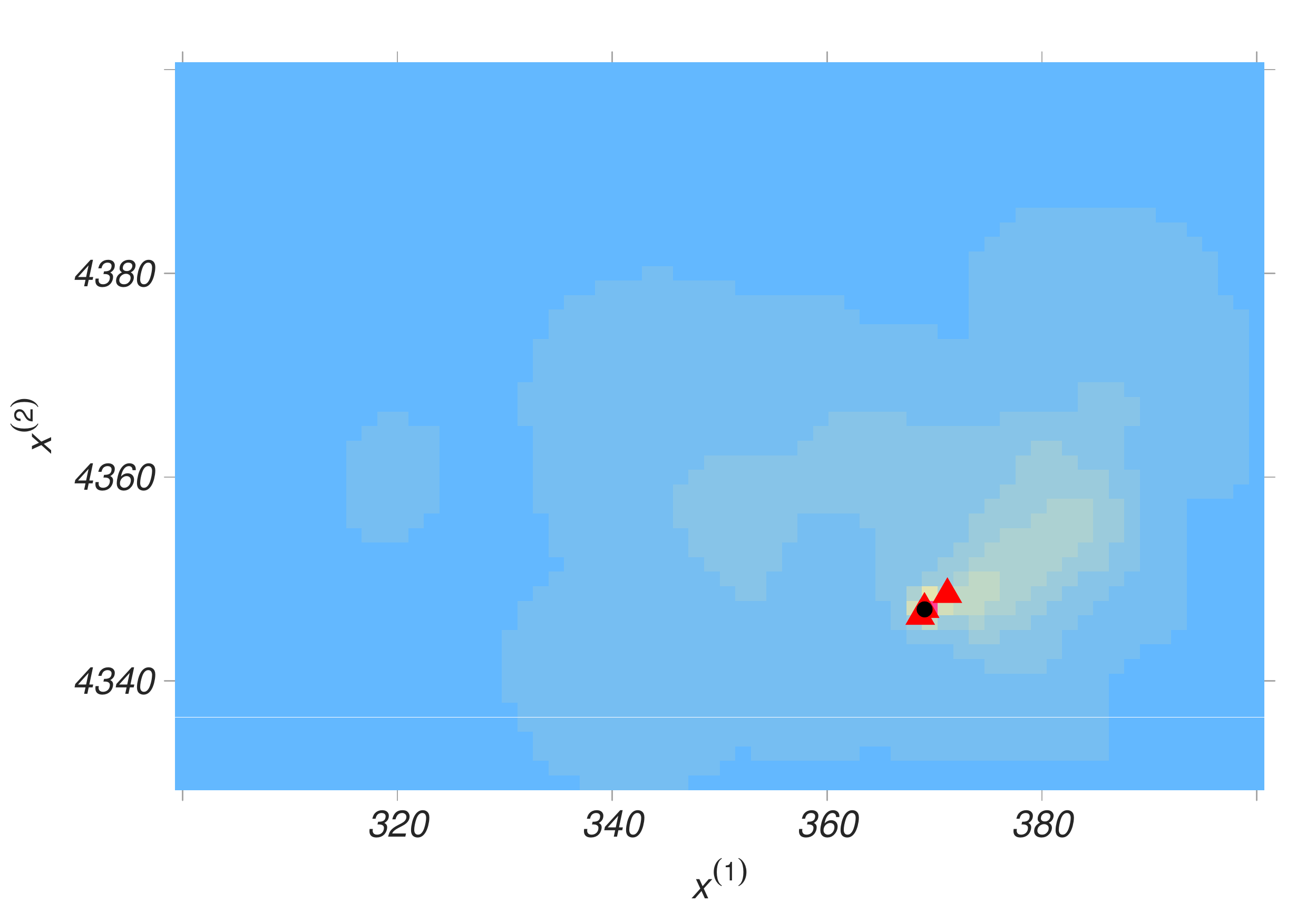}
                 \caption{\textit{Method \hyperref[2ai]{2ai} for a resident without a buffer zone.}  }
                 \label{O:pachMar2ai}
         \end{subfigure}~
    \begin{subfigure}[b]{0.48\textwidth}
                 \includegraphics[trim= 0mm 0mm 0mm  0mm,width=\textwidth]{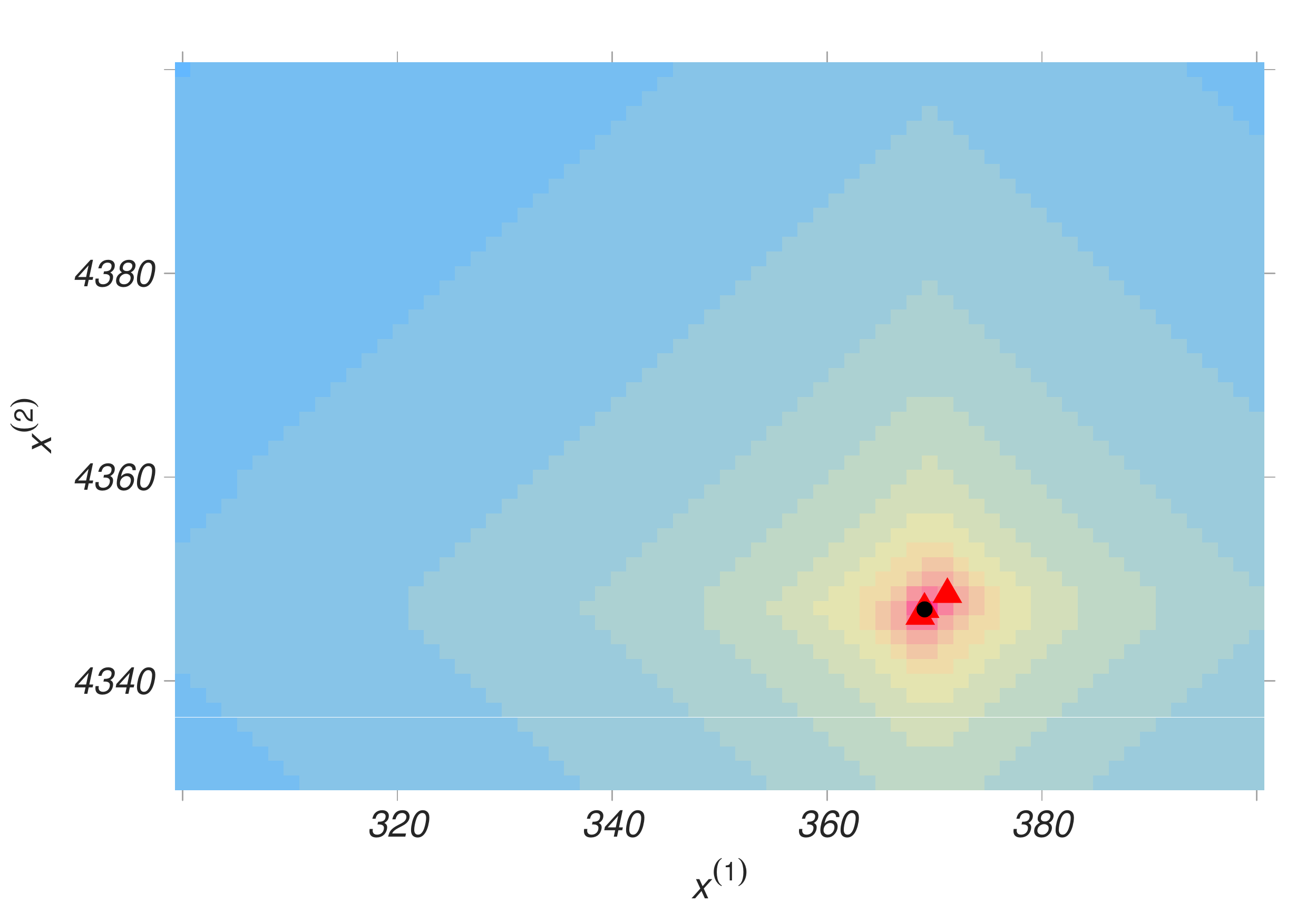}
                 \caption{\textit{Rossmo's model for a resident without a buffer zone.}}
                 \label{O:pachMarRoss}
         \end{subfigure}
         
             \begin{subfigure}[b]{0.48\textwidth}
                 \includegraphics[trim= 0mm 0mm 0mm  0mm,width=\textwidth]{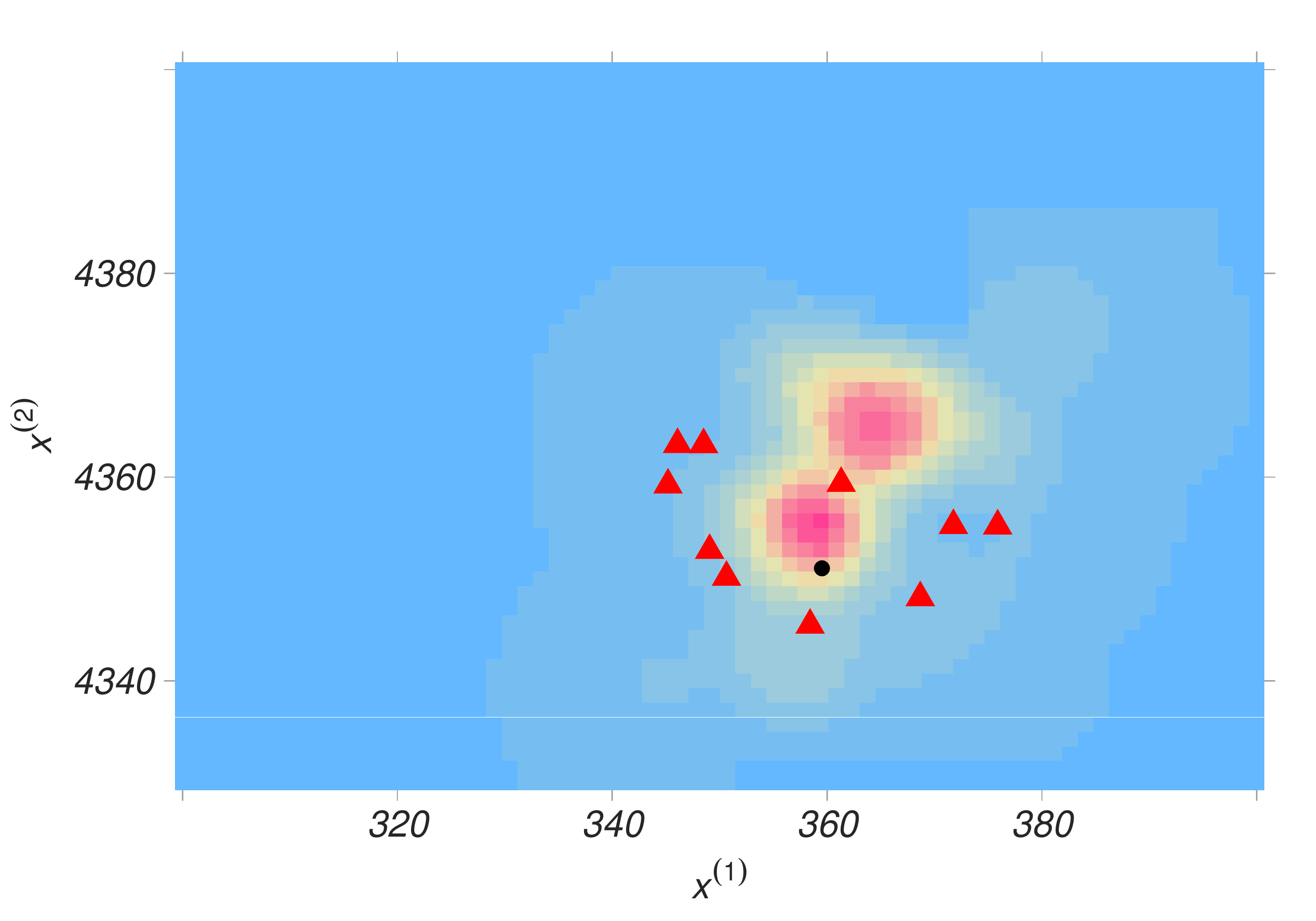}
                 \caption{\textit{Method \hyperref[2ai]{2ai} for a resident with a buffer zone.\\}}
                 \label{O:pachMar2ai2}
         \end{subfigure}~
    \begin{subfigure}[b]{0.48\textwidth}
                 \includegraphics[trim= 0mm 0mm 0mm  0mm,width=\textwidth]{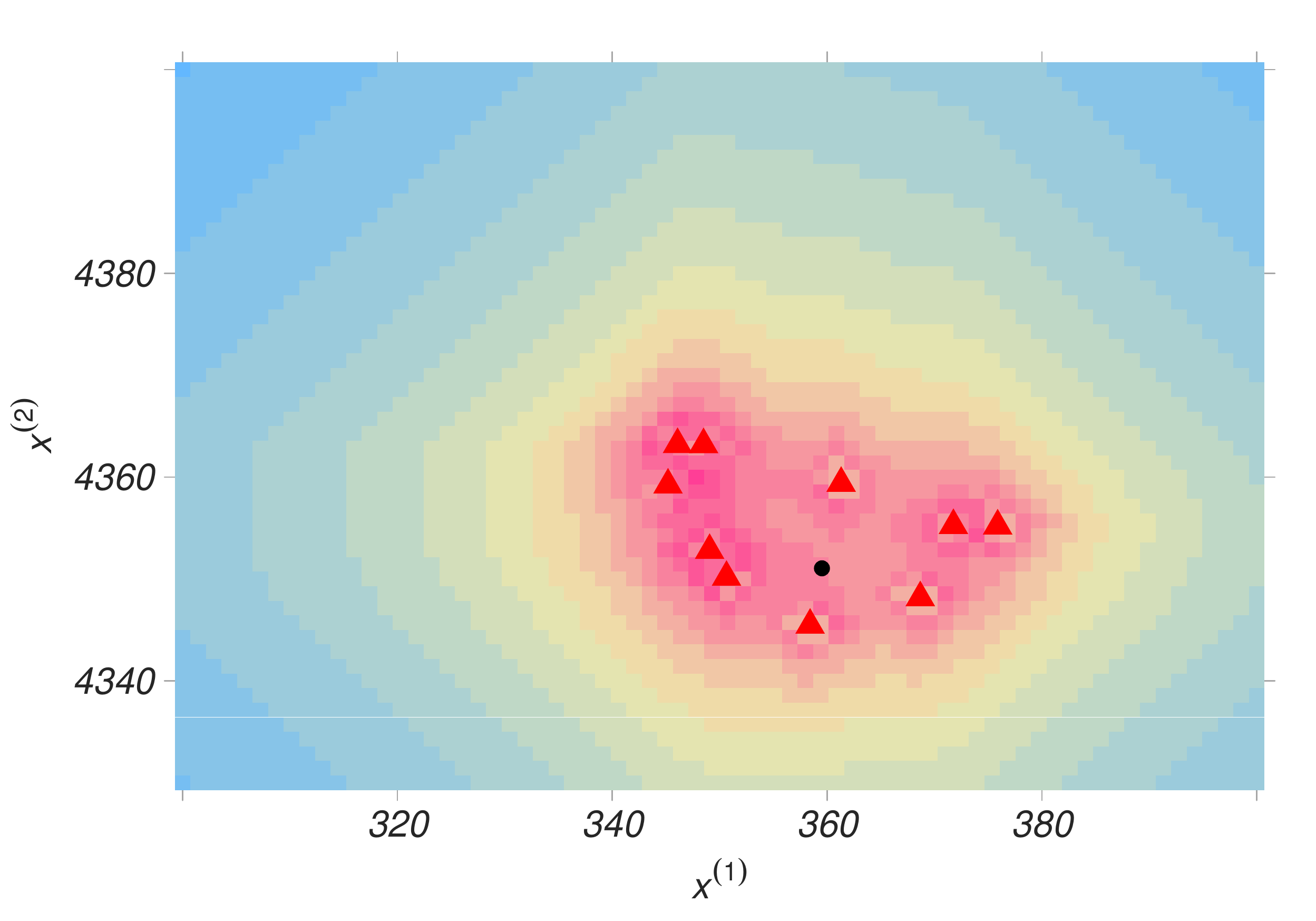}
                 \caption{\textit{Rossmo's model for a resident with a buffer zone.}}
                 \label{O:pachMarRoss2}
         \end{subfigure}
         
             \begin{subfigure}[b]{0.48\textwidth}
                 \includegraphics[trim= 0mm 0mm 0mm  0mm,width=\textwidth]{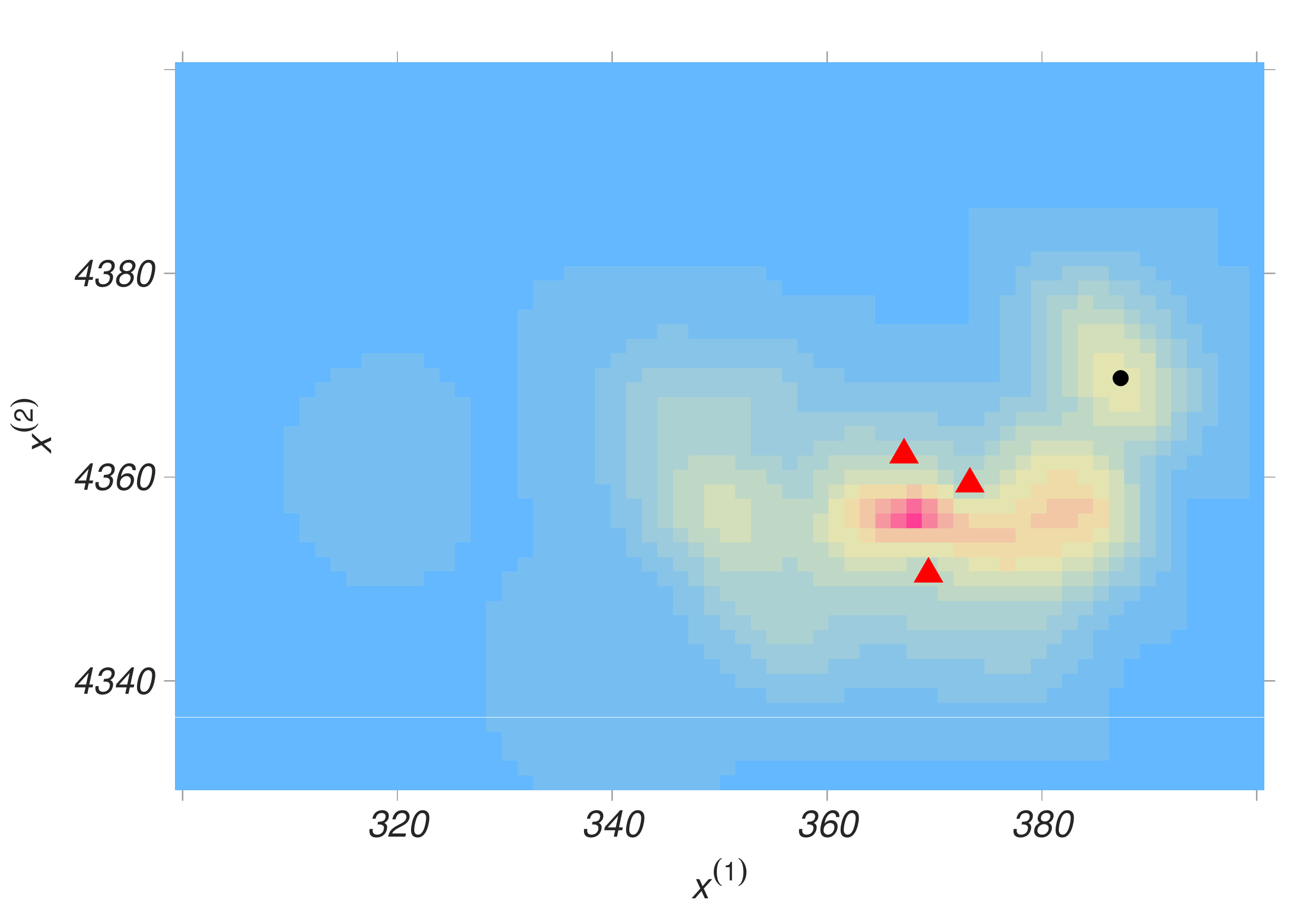}
                 \caption{\textit{Method \hyperref[2ai]{2ai} for a non-resident.}}
                 \label{O:pachCom2ai}
         \end{subfigure}~
    \begin{subfigure}[b]{0.48\textwidth}
                 \includegraphics[trim= 0mm 0mm 0mm  0mm,width=\textwidth]{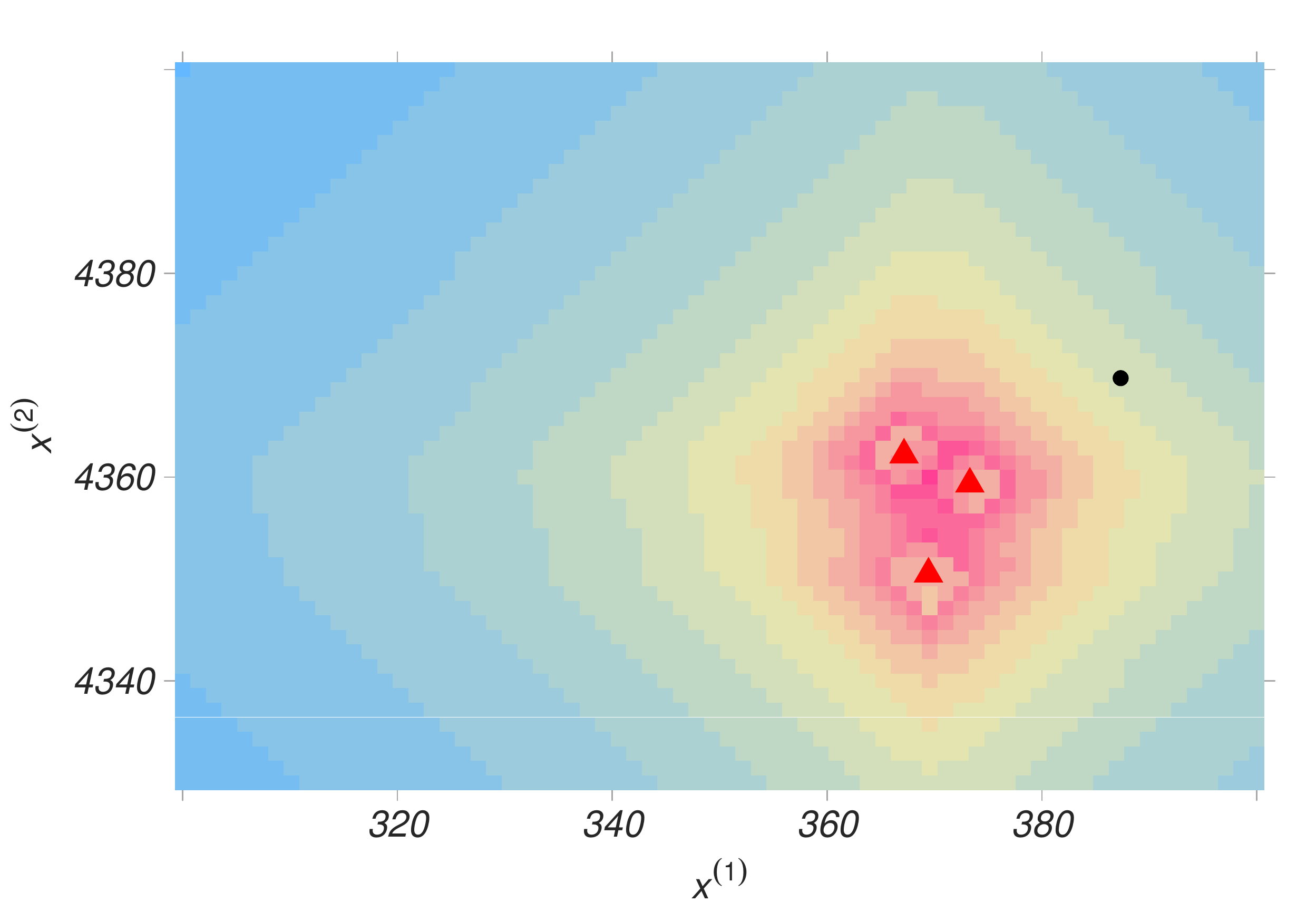}
                 \caption{\textit{Rossmo's model for a non-resident.}}
                 \label{O:pachComRoss}
         \end{subfigure}
\caption{\textit{Level plots indicating how likely is that the area contains the anchor point of the offender (regions with the highest probability are pink, areas with the lowest probability are blue). The red triangles denote crime sites of the offender, the black circle indicates his anchor point.} }
      \label{O:ModelyPachateleData}
    \end{center}
  \end{figure}

 In Fig. \ref{O:ModelyPachateleData}, we can see estimates of the offender’s anchor point -- we use the method \hyperref[2ai]{2ai} and  Rossmo’s approach for each case.  Both methods are similarly  effective for residents without a buffer zone  (Fig. \ref{O:pachMar2ai} and~\ref{O:pachMarRoss}). However, in the case of residents with a buffer zone   (Fig. \ref{O:pachMar2ai2} and~\ref{O:pachMarRoss2}) and also  in the case of non-residents  (Fig. \ref{O:pachCom2ai} and~\ref{O:pachComRoss}), the hit score function using Rossmo’s distance decay function still assumes that the anchor point lies close to  any of the offender’s crime sites. Conversely, the method \hyperref[2ai]{2ai} admits the possibility that the criminal’s anchor point is located at a greater distance from his crime sites.

\section{Conclusion}
We offered  a more complex approach to treating the problem of geographic profiling.
After analyzing the data set on serial criminals from a certain area, we can observe similar tendencies for choosing the 
crime site. Based on this,  we classify the criminals into different types. It often turns out that a similar distribution of crime sites 
corresponds to a very similar behaviour when choosing the crime site.   
Thanks to this we can, for  a given criminal,  construct a model based on the  behavior of 
criminals of the same type.

We divided the offenders from our data set into two basic groups -- residents and non-residents. This distinction is very similar to the marauder and the commuter hypothesis. The main idea of the overlap of criminal and anchor areas is the same. However, the models which are able to describe the behavior of commuters (in our case non-residents) are very rare in the literature. 
One of the few is suggested in the paper of \citet{mohler2}. Their approach, based on solving a stochastic 
differential equation, leads to a model which for a suitable choice of parameters characterizes  the 
nature of commuters (non-residents) well. In this paper, we offered a simpler, purely Bayesian model, which preserves the main features and flexibility of models suggested in \citet{mohler2}.

The considered models  capture the criminal's behavior better than the models of \citet{oleary1}, since our setup  is based on data from similarly 
behaving criminals. At the same time, they are easier to apply and interpret than the models of \textit{Mohler} and \textit{Short}, while 
keeping the ability to cover a broad spectrum of offenders. Further, the paper showed how to distinguish the types of  offenders based on the spatial distribution of their crime sites.

In case we were not able to assign the given criminal to a certain group of already investigated criminals, 
it was natural to use multimodel inference. It allows to incorporate  all considered possibilities into the model.
In particular, this approach was used to deal with the case when the criminal can be a resident or a non-resident. In our case it turned out that it is suitable to derive weights for multimodel inference by frequencies of residents and non-residents in the dataset.

We used the standard Rossmo's model as a benchmark. Our model substantially overperformed the benchmark, 
especially in the presence of non-residents.

\clearpage

\bibliographystyle{apacite}
\bibliography{citace}

\begin{thebibliography}{}

\bibitem [\protect \citeauthoryear {%
Bolstad%
}{%
Bolstad%
}{%
{\protect \APACyear {c2007}}%
}]{%
bolstad}
\APACinsertmetastar {%
bolstad}%
\begin{APACrefauthors}%
Bolstad, W\BPBI M.%
\end{APACrefauthors}%
\unskip\
\newblock
\APACrefYear{c2007}.
\newblock
\APACrefbtitle {Introduction to Bayesian statistics} {Introduction to bayesian
  statistics}\ (\PrintOrdinal{2nd ed.}\ \BEd).
\newblock
\APACaddressPublisher{Hoboken, N.J.}{John Wiley}.
\PrintBackRefs{\CurrentBib}

\bibitem [\protect \citeauthoryear {%
Brantingham%
\ \BBA {} Brantingham%
}{%
Brantingham%
\ \BBA {} Brantingham%
}{%
{\protect \APACyear {1993}}%
}]{%
brantingham}
\APACinsertmetastar {%
brantingham}%
\begin{APACrefauthors}%
Brantingham, P\BPBI L.%
\BCBT {}\ \BBA {} Brantingham, P\BPBI J.%
\end{APACrefauthors}%
\unskip\
\newblock
\APACrefYearMonthDay{1993}{}{}.
\newblock
{\BBOQ}\APACrefatitle {Nodes, paths and edges} {Nodes, paths and edges}.{\BBCQ}
\newblock
\APACjournalVolNumPages{Journal of Environmental Psychology}{13}{1}{3--28}.
\newblock
\begin{APACrefURL}
  \url{http://linkinghub.elsevier.com/retrieve/pii/S0272494405802129}
  \end{APACrefURL}
\PrintBackRefs{\CurrentBib}

\bibitem [\protect \citeauthoryear {%
Burnham%
, Anderson%
\BCBL {}\ \BBA {} Burnham%
}{%
Burnham%
\ \protect \BOthers {.}}{%
{\protect \APACyear {c2002}}%
}]{%
burnham}
\APACinsertmetastar {%
burnham}%
\begin{APACrefauthors}%
Burnham, K\BPBI P.%
, Anderson, D\BPBI R.%
\BCBL {}\ \BBA {} Burnham, K\BPBI P.%
\end{APACrefauthors}%
\unskip\
\newblock
\APACrefYear{c2002}.
\newblock
\APACrefbtitle {Model selection and multimodel inference} {Model selection and
  multimodel inference}\ (\PrintOrdinal{2nd ed.}\ \BEd).
\newblock
\APACaddressPublisher{New York}{Springer}.
\PrintBackRefs{\CurrentBib}

\bibitem [\protect \citeauthoryear {%
Canter%
}{%
Canter%
}{%
{\protect \APACyear {1996}}%
}]{%
canter}
\APACinsertmetastar {%
canter}%
\begin{APACrefauthors}%
Canter, D.%
\end{APACrefauthors}%
\unskip\
\newblock
\APACrefYear{1996}.
\newblock
\APACrefbtitle {Psychology in action} {Psychology in action}.
\newblock
\APACaddressPublisher{Aldershot, England Brookfield, Vt., USA}{Dartmouth}.
\PrintBackRefs{\CurrentBib}

\bibitem [\protect \citeauthoryear {%
Canter%
, Coffey%
, Huntley%
\BCBL {}\ \BBA {} Missen%
}{%
Canter%
\ \protect \BOthers {.}}{%
{\protect \APACyear {2000}}%
}]{%
canter2}
\APACinsertmetastar {%
canter2}%
\begin{APACrefauthors}%
Canter, D.%
, Coffey, T.%
, Huntley, M.%
\BCBL {}\ \BBA {} Missen, C.%
\end{APACrefauthors}%
\unskip\
\newblock
\APACrefYearMonthDay{2000}{}{}.
\newblock
{\BBOQ}\APACrefatitle {Predicting Serial Killers' Home Base Using a Decision
  Support System} {Predicting serial killers' home base using a decision
  support system}.{\BBCQ}
\newblock
\APACjournalVolNumPages{Journal of Quantitative Criminology}{16}{4}{457--478}.
\newblock
\begin{APACrefURL} \url{http://dx.doi.org/10.1023/A%3A1007551316253}
  \end{APACrefURL}
\newblock
\begin{APACrefDOI} \doi{10.1023/A:1007551316253} \end{APACrefDOI}
\PrintBackRefs{\CurrentBib}

\bibitem [\protect \citeauthoryear {%
Carlin%
}{%
Carlin%
}{%
{\protect \APACyear {2000}}%
}]{%
carlin}
\APACinsertmetastar {%
carlin}%
\begin{APACrefauthors}%
Carlin, B.%
\end{APACrefauthors}%
\unskip\
\newblock
\APACrefYear{2000}.
\newblock
\APACrefbtitle {Bayes and Empirical Bayes methods for data analysis} {Bayes and
  empirical bayes methods for data analysis}.
\newblock
\APACaddressPublisher{Boca Raton}{Chapman \& Hall/CRC}.
\PrintBackRefs{\CurrentBib}

\bibitem [\protect \citeauthoryear {%
Damien%
, Dellaportas%
, Polson%
, Stephens%
\BCBL {}\ \BBA {} Smith%
}{%
Damien%
\ \protect \BOthers {.}}{%
{\protect \APACyear {2013}}%
}]{%
damien}
\APACinsertmetastar {%
damien}%
\begin{APACrefauthors}%
Damien, P.%
, Dellaportas, P.%
, Polson, N\BPBI G.%
, Stephens, D\BPBI A.%
\BCBL {}\ \BBA {} Smith, A\BPBI F.%
\end{APACrefauthors}%
\unskip\
\newblock
\APACrefYear{2013}.
\newblock
\APACrefbtitle {Bayesian theory and applications} {Bayesian theory and
  applications}\ (\PrintOrdinal{First edition.}\ \BEd).
\newblock
\APACaddressPublisher{Oxford}{Oxford University Press}.
\PrintBackRefs{\CurrentBib}

\bibitem [\protect \citeauthoryear {%
Kawase%
}{%
Kawase%
}{%
{\protect \APACyear {2011}}%
}]{%
kawase1}
\APACinsertmetastar {%
kawase1}%
\begin{APACrefauthors}%
Kawase, K.%
\end{APACrefauthors}%
\unskip\
\newblock
\APACrefYearMonthDay{2011}{}{}.
\newblock
{\BBOQ}\APACrefatitle {A General Formula for Calculating Meridian Arc Length
  and its Application to Coordinate Conversion in the Gauss-Krüger Projection}
  {A general formula for calculating meridian arc length and its application to
  coordinate conversion in the gauss-krüger projection}.{\BBCQ}
\newblock
\APACjournalVolNumPages{Bulletin of the Geospatial Information Authority of
  Japan}{59}{}{1--13}.
\newblock
\begin{APACrefURL} \url{http://www.gsi.go.jp/common/000062452.pdf}
  \end{APACrefURL}
\PrintBackRefs{\CurrentBib}

\bibitem [\protect \citeauthoryear {%
Kawase%
}{%
Kawase%
}{%
{\protect \APACyear {2012}}%
}]{%
kawase2}
\APACinsertmetastar {%
kawase2}%
\begin{APACrefauthors}%
Kawase, K.%
\end{APACrefauthors}%
\unskip\
\newblock
\APACrefYearMonthDay{2012}{}{}.
\newblock
{\BBOQ}\APACrefatitle {The environmental range of serial rapists} {The
  environmental range of serial rapists}.{\BBCQ}
\newblock
\APACjournalVolNumPages{Bulletin of the Geospatial Information Authority of
  Japan}{60}{}{1--6}.
\newblock
\begin{APACrefURL} \url{http://www.gsi.go.jp/common/000065826.pdf}
  \end{APACrefURL}
\PrintBackRefs{\CurrentBib}

\bibitem [\protect \citeauthoryear {%
Kennedy%
}{%
Kennedy%
}{%
{\protect \APACyear {2000}}%
}]{%
kennedy}
\APACinsertmetastar {%
kennedy}%
\begin{APACrefauthors}%
Kennedy, M.%
\end{APACrefauthors}%
\unskip\
\newblock
\APACrefYear{2000}.
\newblock
\APACrefbtitle {Understanding map projections : GIS by ESRI} {Understanding map
  projections : Gis by esri}.
\newblock
\APACaddressPublisher{Redlands, CA}{ESRI}.
\PrintBackRefs{\CurrentBib}

\bibitem [\protect \citeauthoryear {%
Kooperberg%
\ \BBA {} Stone%
}{%
Kooperberg%
\ \BBA {} Stone%
}{%
{\protect \APACyear {1991}}%
}]{%
kooperberg}
\APACinsertmetastar {%
kooperberg}%
\begin{APACrefauthors}%
Kooperberg, C.%
\BCBT {}\ \BBA {} Stone, C\BPBI J.%
\end{APACrefauthors}%
\unskip\
\newblock
\APACrefYearMonthDay{1991}{}{}.
\newblock
{\BBOQ}\APACrefatitle {A study of logspline density estimation} {A study of
  logspline density estimation}.{\BBCQ}
\newblock
\APACjournalVolNumPages{Computational Statistics \& Data
  Analysis}{12}{3}{327--347}.
\newblock
\begin{APACrefURL}
  \url{http://www.sciencedirect.com/science/article/pii/016794739190115I}
  \end{APACrefURL}
\newblock
\begin{APACrefDOI} \doi{http://dx.doi.org/10.1016/0167-9473(91)90115-I}
  \end{APACrefDOI}
\PrintBackRefs{\CurrentBib}

\bibitem [\protect \citeauthoryear {%
Levine%
}{%
Levine%
}{%
{\protect \APACyear {2008}}%
}]{%
levine}
\APACinsertmetastar {%
levine}%
\begin{APACrefauthors}%
Levine, N.%
\end{APACrefauthors}%
\unskip\
\newblock
\APACrefYearMonthDay{2008}{}{}.
\newblock
{\BBOQ}\APACrefatitle {CrimeStat: A Spatial Statistical Program for the
  Analysis of Crime Incidents} {Crimestat: A spatial statistical program for
  the analysis of crime incidents}.{\BBCQ}
\newblock
\BIn{} \APACrefbtitle {Encyclopedia of GIS} {Encyclopedia of gis}\ (\BPGS\
  187--193).
\newblock
\APACaddressPublisher{}{Springer US}.
\newblock
\begin{APACrefURL} \url{http://dx.doi.org/10.1007/978-0-387-35973-1_229}
  \end{APACrefURL}
\newblock
\begin{APACrefDOI} \doi{10.1007/978-0-387-35973-1_229} \end{APACrefDOI}
\PrintBackRefs{\CurrentBib}

\bibitem [\protect \citeauthoryear {%
Mohler%
\ \BBA {} Short%
}{%
Mohler%
\ \BBA {} Short%
}{%
{\protect \APACyear {2012}}%
}]{%
mohler2}
\APACinsertmetastar {%
mohler2}%
\begin{APACrefauthors}%
Mohler, G\BPBI O.%
\BCBT {}\ \BBA {} Short, M\BPBI B.%
\end{APACrefauthors}%
\unskip\
\newblock
\APACrefYearMonthDay{2012}{}{}.
\newblock
{\BBOQ}\APACrefatitle {Geographic Profiling from Kinetic Models of Criminal
  Behavior} {Geographic profiling from kinetic models of criminal
  behavior}.{\BBCQ}
\newblock
\APACjournalVolNumPages{SIAM Journal on Applied Mathematics}{72}{1}{163--180}.
\newblock
\begin{APACrefURL} \url{http://dx.doi.org/10.1137/100794080} \end{APACrefURL}
\newblock
\begin{APACrefDOI} \doi{10.1137/100794080} \end{APACrefDOI}
\PrintBackRefs{\CurrentBib}

\bibitem [\protect \citeauthoryear {%
O'Leary%
}{%
O'Leary%
}{%
{\protect \APACyear {2009}}%
{\protect \APACexlab {{\protect \BCnt {1}}}}}]{%
oleary1}
\APACinsertmetastar {%
oleary1}%
\begin{APACrefauthors}%
O'Leary, M.%
\end{APACrefauthors}%
\unskip\
\newblock
\APACrefYearMonthDay{2009{\protect \BCnt {1}}}{}{}.
\newblock
{\BBOQ}\APACrefatitle {The mathematics of geographic profiling} {The
  mathematics of geographic profiling}.{\BBCQ}
\newblock
\APACjournalVolNumPages{Journal of Investigative Psychology and Offender
  Profiling}{6}{3}{253--265}.
\newblock
\begin{APACrefURL} \url{http://dx.doi.org/10.1002/jip.111} \end{APACrefURL}
\newblock
\begin{APACrefDOI} \doi{10.1002/jip.111} \end{APACrefDOI}
\PrintBackRefs{\CurrentBib}

\bibitem [\protect \citeauthoryear {%
O'Leary%
}{%
O'Leary%
}{%
{\protect \APACyear {2009}}%
{\protect \APACexlab {{\protect \BCnt {2}}}}}]{%
oleary2}
\APACinsertmetastar {%
oleary2}%
\begin{APACrefauthors}%
O'Leary, M.%
\end{APACrefauthors}%
\unskip\
\newblock
\APACrefYearMonthDay{2009{\protect \BCnt {2}}}{}{}.
\newblock
{\BBOQ}\APACrefatitle {A new mathematical approach to geographic profiling} {A
  new mathematical approach to geographic profiling}.{\BBCQ}
\newblock
\APACjournalVolNumPages{Towson University}{}{}{}.
\PrintBackRefs{\CurrentBib}

\bibitem [\protect \citeauthoryear {%
O'Leary%
}{%
O'Leary%
}{%
{\protect \APACyear {{2010}}}%
{\protect \APACexlab {{\protect \BCnt {1}}}}}]{%
oleary5}
\APACinsertmetastar {%
oleary5}%
\begin{APACrefauthors}%
O'Leary, M.%
\end{APACrefauthors}%
\unskip\
\newblock
\APACrefYearMonthDay{{2010}{\protect \BCnt {1}}}{}{}.
\newblock
{\BBOQ}\APACrefatitle {Implementing a Bayesian Approach to Criminal Geographic
  Profiling} {Implementing a bayesian approach to criminal geographic
  profiling}.{\BBCQ}
\newblock
\BIn{} \APACrefbtitle {Proceedings of the 1st International Conference and
  Exhibition on Computing for Geospatial Research \& Application} {Proceedings
  of the 1st international conference and exhibition on computing for
  geospatial research \& application}\ (\BPGS\ 59:1--59:8).
\newblock
\APACaddressPublisher{New York, NY, USA}{ACM}.
\newblock
\begin{APACrefURL} \url{http://doi.acm.org/10.1145/1823854.1823920}
  \end{APACrefURL}
\newblock
\begin{APACrefDOI} \doi{10.1145/1823854.1823920} \end{APACrefDOI}
\PrintBackRefs{\CurrentBib}

\bibitem [\protect \citeauthoryear {%
O'Leary%
}{%
O'Leary%
}{%
{\protect \APACyear {{2010}}}%
{\protect \APACexlab {{\protect \BCnt {2}}}}}]{%
oleary3}
\APACinsertmetastar {%
oleary3}%
\begin{APACrefauthors}%
O'Leary, M.%
\end{APACrefauthors}%
\unskip\
\newblock
\APACrefYearMonthDay{{2010}{\protect \BCnt {2}}}{}{}.
\newblock
{\BBOQ}\APACrefatitle {Multimodel Inference and Geographic Profiling}
  {Multimodel inference and geographic profiling}.{\BBCQ}
\newblock
\APACjournalVolNumPages{Crime Mapping}{2}{1}{}.
\PrintBackRefs{\CurrentBib}

\bibitem [\protect \citeauthoryear {%
Robert%
}{%
Robert%
}{%
{\protect \APACyear {2007}}%
}]{%
robert}
\APACinsertmetastar {%
robert}%
\begin{APACrefauthors}%
Robert, C.%
\end{APACrefauthors}%
\unskip\
\newblock
\APACrefYear{2007}.
\newblock
\APACrefbtitle {The Bayesian choice from decision-theoretic foundations to
  computational implementation} {The bayesian choice from decision-theoretic
  foundations to computational implementation}.
\newblock
\APACaddressPublisher{New York}{Springer}.
\PrintBackRefs{\CurrentBib}

\bibitem [\protect \citeauthoryear {%
D.~Rossmo%
}{%
D.~Rossmo%
}{%
{\protect \APACyear {2000}}%
}]{%
rossmo1}
\APACinsertmetastar {%
rossmo1}%
\begin{APACrefauthors}%
Rossmo, D.%
\end{APACrefauthors}%
\unskip\
\newblock
\APACrefYear{2000}.
\newblock
\APACrefbtitle {Geographic profiling} {Geographic profiling}.
\newblock
\APACaddressPublisher{Boca Raton, Fla}{CRC Press}.
\PrintBackRefs{\CurrentBib}

\bibitem [\protect \citeauthoryear {%
D\BPBI K.~Rossmo%
}{%
D\BPBI K.~Rossmo%
}{%
{\protect \APACyear {1995}}%
}]{%
rossmo2}
\APACinsertmetastar {%
rossmo2}%
\begin{APACrefauthors}%
Rossmo, D\BPBI K.%
\end{APACrefauthors}%
\unskip\
\newblock
\APACrefYear{1995}.
\unskip\
\newblock
\APACrefbtitle {Geographic profiling: Target patterns of serial murderers}
  {Geographic profiling: Target patterns of serial murderers}\
  \APACtypeAddressSchool {\BUPhD}{}{}.
\unskip\
\newblock
\APACaddressSchool {}{Theses (School of Criminology)/Simon Fraser University}.
\PrintBackRefs{\CurrentBib}

\bibitem [\protect \citeauthoryear {%
Sivia%
}{%
Sivia%
}{%
{\protect \APACyear {2006}}%
}]{%
sivia}
\APACinsertmetastar {%
sivia}%
\begin{APACrefauthors}%
Sivia, D\BPBI S.%
\end{APACrefauthors}%
\unskip\
\newblock
\APACrefYear{2006}.
\newblock
\APACrefbtitle {Data analysis a Bayesian tutorial} {Data analysis a bayesian
  tutorial}.
\newblock
\APACaddressPublisher{Oxford New York}{Oxford University Press}.
\PrintBackRefs{\CurrentBib}

\end{thebibliography}

\end{document}